\documentclass[traditabstract]{aa}
\usepackage{natbib}
\bibpunct{(}{)}{;}{a}{}{,}
\usepackage{graphicx}
\usepackage{amssymb}
\usepackage{amsmath}
\usepackage{url}
\usepackage{txfonts}
\usepackage{rotating}
\usepackage{xspace}

\newcommand{\Cyg}{\mbox{Cyg\,X-1}\xspace}
\newcommand{\RXTE}{\textsl{RXTE}\xspace}

\newcommand\figHid{
 \begin{figure*}
  \includegraphics[width=\columnwidth]{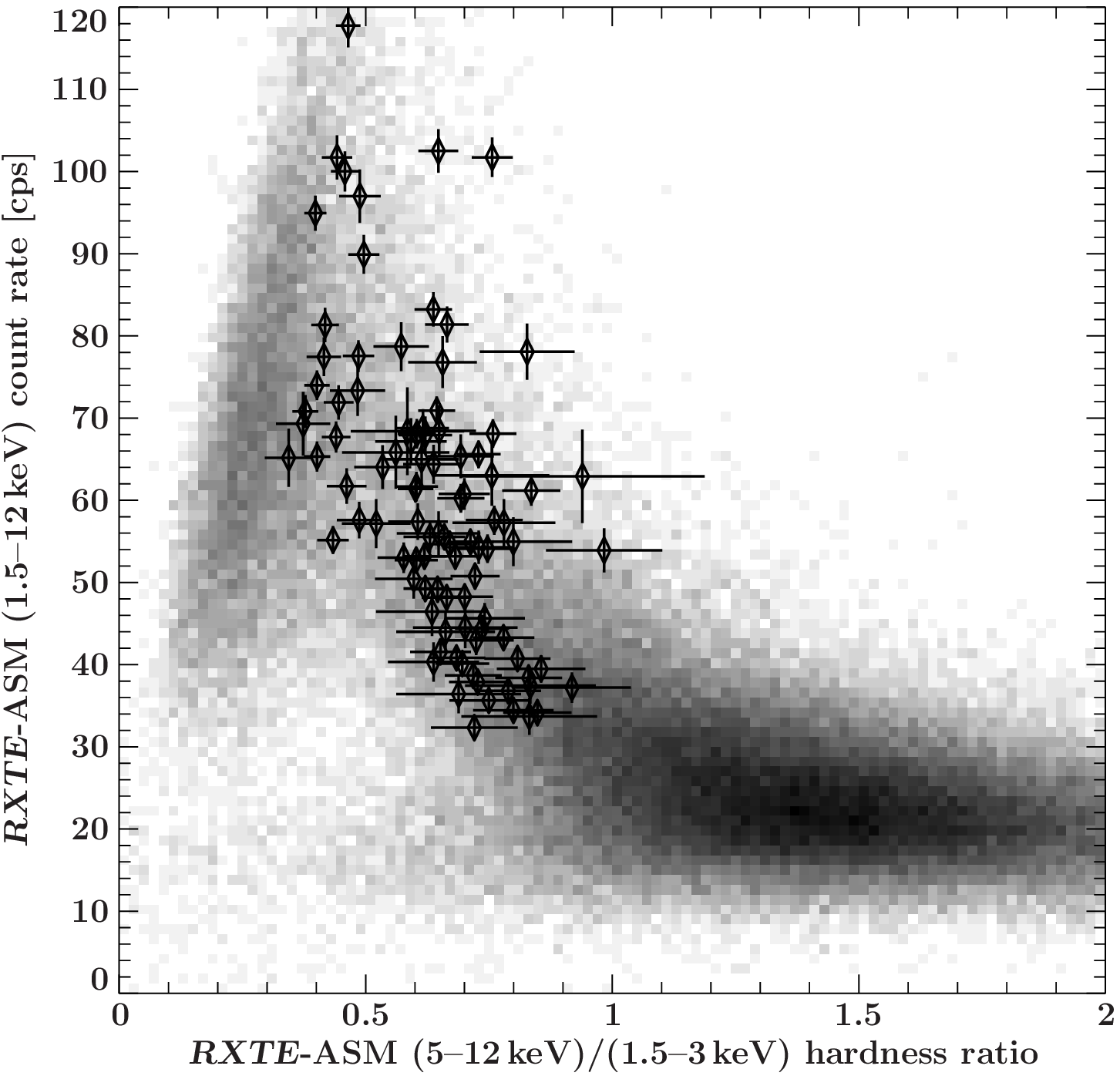}\hfill
  \includegraphics[width=\columnwidth]{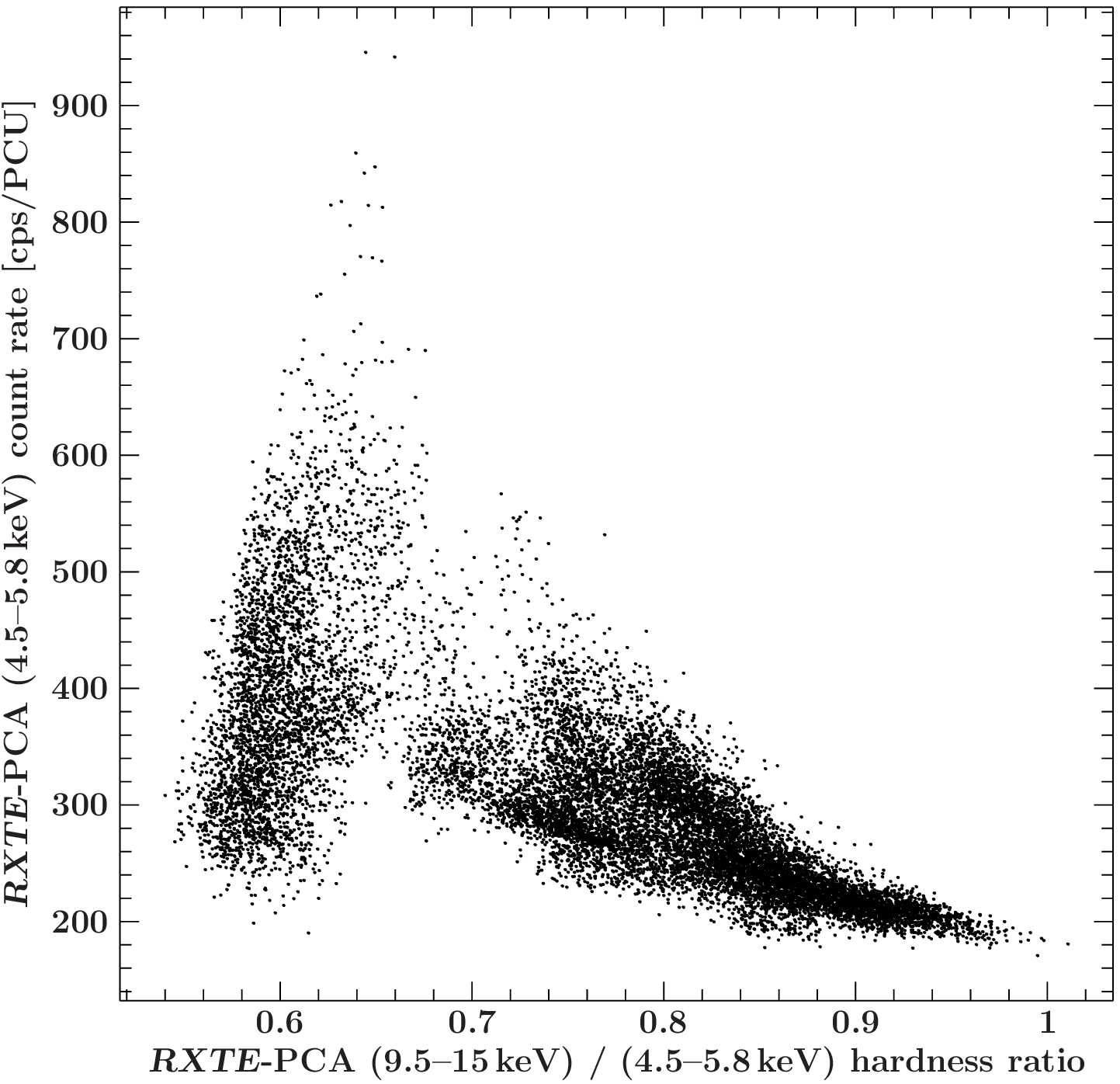}
  \caption{\emph{Left}: Hardness-intensity-diagram of \Cyg using
    \textsl{RXTE}-ASM data. The hardness is defined here as the ratio of
    the 5--12\,keV and the 1.5--3\,keV ASM count rates; the intensity is
    the total 1.5--12\,keV ASM count rate \citep[see also][their
    Fig.~5]{Fender2006}. The grayscale represents the single-dwell
    (90\,s) ASM-data from 1996~January until 2010~August. Their
    characteristic density forms two regions in the HID, indicating two
    different states. The data analyzed in this work (black symbols)
    cover the transition region. \emph{Right:} The transitional nature
    of the observation is confirmed by the HID calculated from 16\,s PCA
    data.
  }
  \label{fig:hid}
 \end{figure*}
}

\newcommand\figMultiSpec{
 \begin{figure*}
  \includegraphics[width=\textwidth]{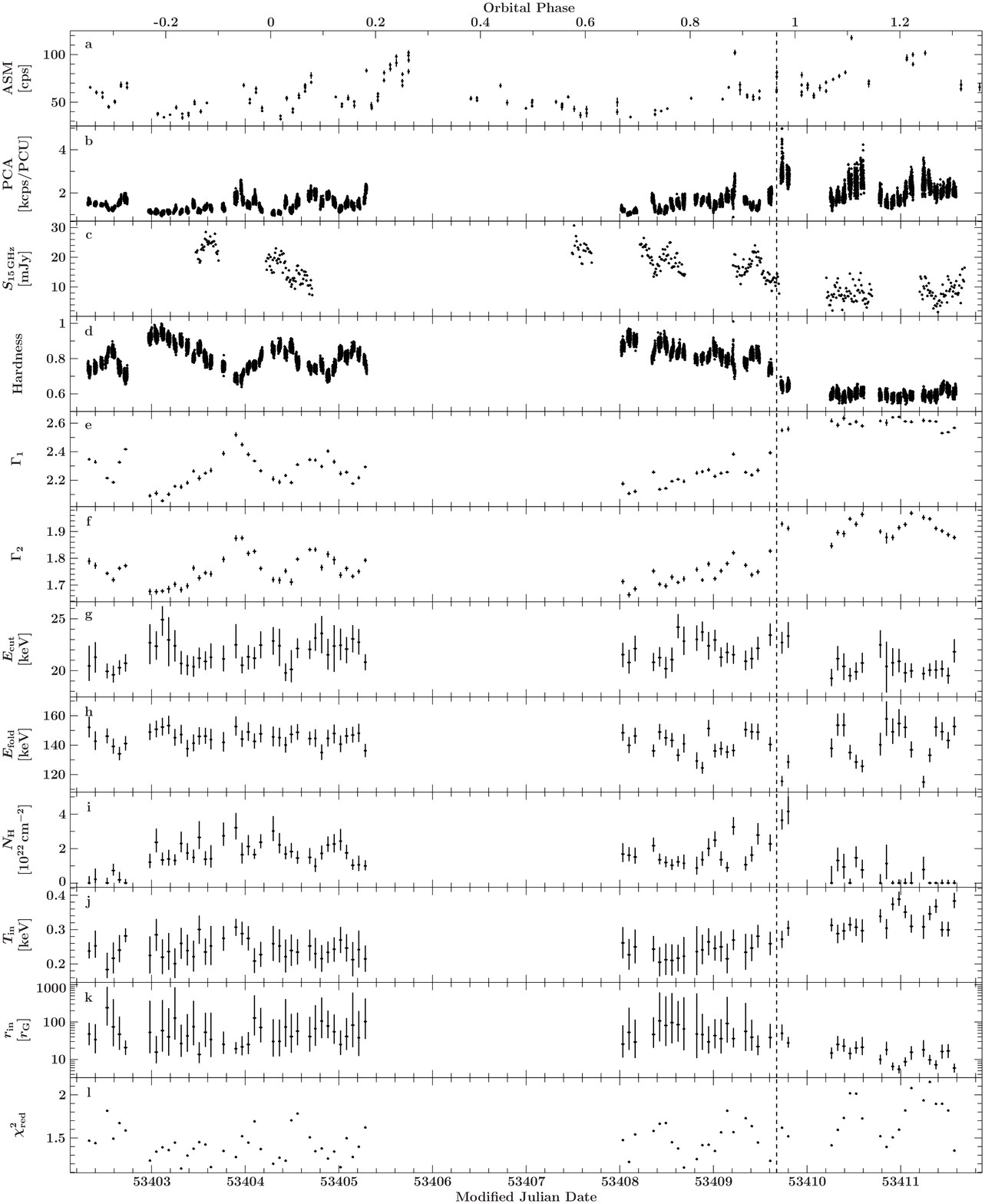}
  \caption{Temporal evolution of the count rates and spectral parameters
    of \Cyg during the observation period.
    The upper three panels are the light curves of \Cyg measured with
    (a) \RXTE-ASM (1.5--12\,keV),
    (b) PCA (total energy range 2--60\,keV) and
    (c) the Ryle Telescope (at 15\,GHz).
    (d) The hardness is the same as in Fig.~\ref{fig:hid_lag},
        namely the ratio of the PCA count rates in the energy bands
	(9.5--15\,keV)/(4.5--5.8\,keV).
    The bottom panels show the temporal evolution of
    (e) the photon index $\Gamma_1$,
    (f) the photon index above the spectral break $\Gamma_2$,
    (g) the cutoff and
    (h) folding energy of the high-energy cutoff,
    (i) the hydrogen column,
    (j) the inner temperature and
    (k) the inner radius of the accretion disk, and
    (l) the reduced $\chi^2$ of the best fits.
    The dashed vertical line indicates the transition.
  }
  \label{fig:multi_spec}
 \end{figure*}
}

\newcommand\figGoneGtwo{
 \begin{figure}
  \includegraphics[width=\columnwidth]{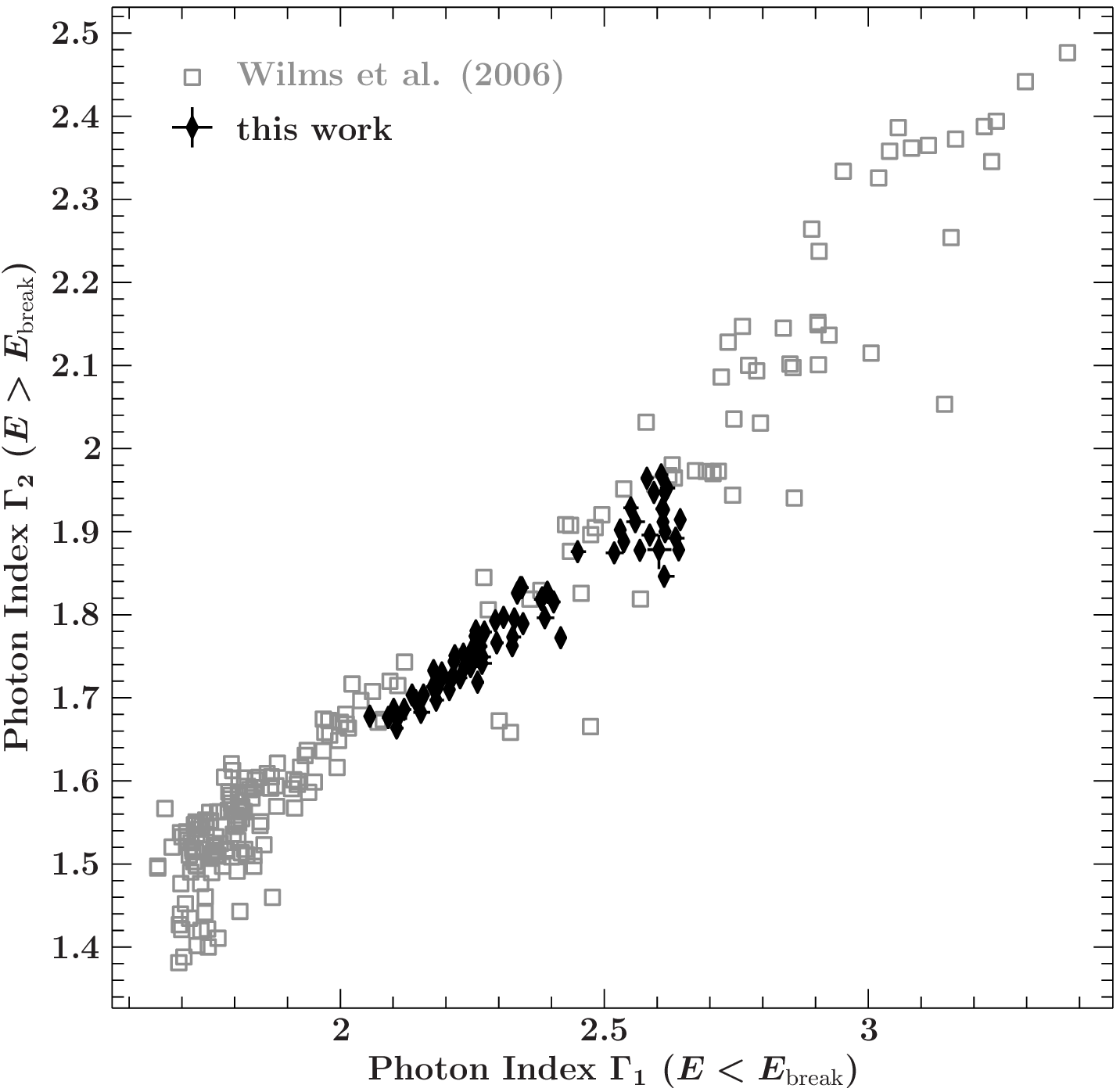}
  \caption{Correlation between the photon indices below and above
    the break energy, derived here for single orbit observations,
    is identical to the linear correlation present in the long-term
    spectral evolution from 1999--2004 \citep{Wilms2006}. Note that this
    correlation extends across hard and soft states.
  }
  \label{fig:G1_G2}
 \end{figure}
}

\newcommand\figGoneRadio{
 \begin{figure}
  \includegraphics[width=\columnwidth]{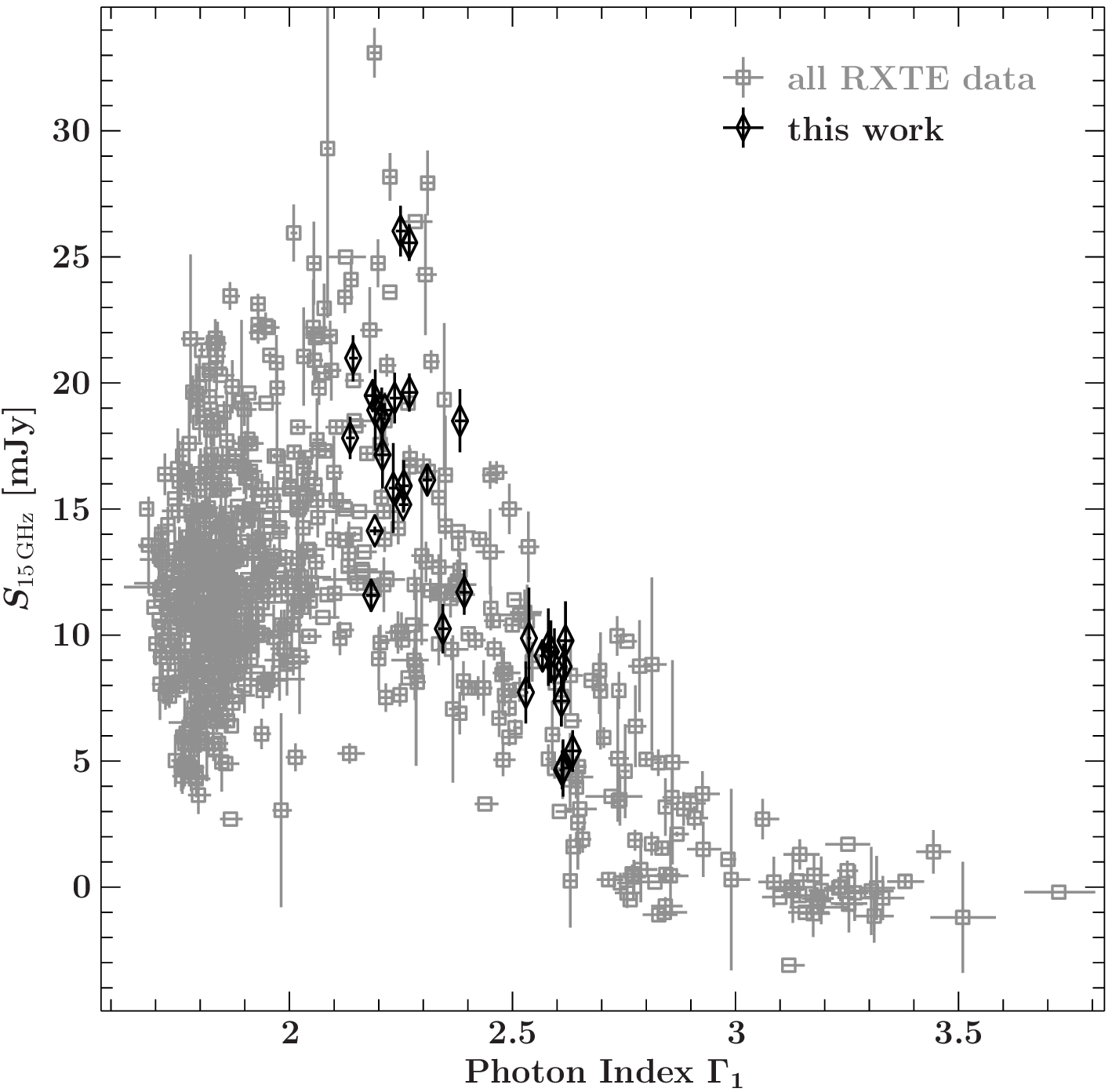}
  \caption{Relation between the 15\,GHz radio flux and the photon
    index $\Gamma_1$.
  }
  \label{fig:g1_radio}
 \end{figure}
}

\newcommand\figPsdSeq{
 \begin{figure*}
  \includegraphics[width=\textwidth]{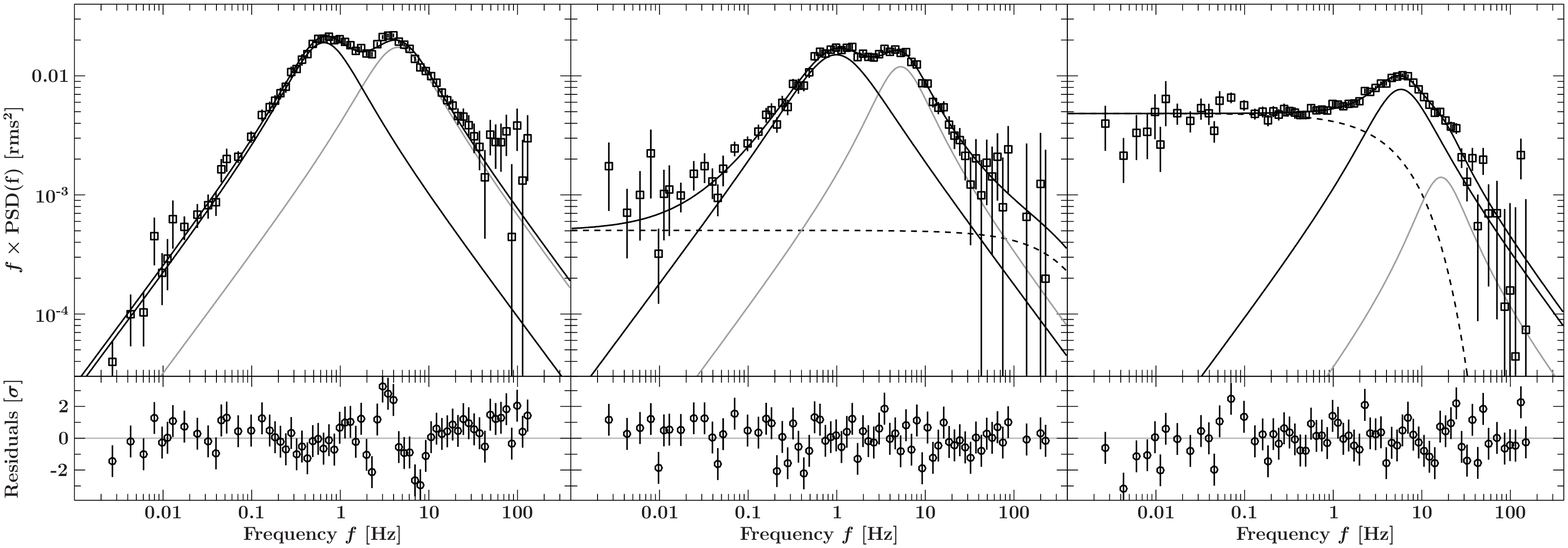}
  \caption{Examples of characteristic power spectra.  The solid gray
    lines show the Lorentzian profiles, and the dashed gray lines a cutoff
    power law \emph{Left:} PSD from an observation with one of the
    hardest spectra in our data set.  $f\times$PSD$(f)$ vanishes for low
    frequencies and shows a double-humped shape, with no requirement for
    a power law in addition to the two Lorentzians.  There is a slight
    indication for a third Lorentzian profile with a peak frequency
    between about 40--80\,Hz.
    \emph{Middle:} PSD obtained from a transitional state,
    requiring an additional power law at low frequencies.
    \emph{Right:} PSD in the soft state,
    with yet stronger contribution of the power law.
    This sequence of power spectra shows that the Lorentzian peaks
    shift to higher frequencies as the X-ray spectrum becomes softer.
  }
  \label{fig:psd_seq}
 \end{figure*}
}

\newcommand\figLoneLtwo{
 \begin{figure}
  \includegraphics[width=\columnwidth]{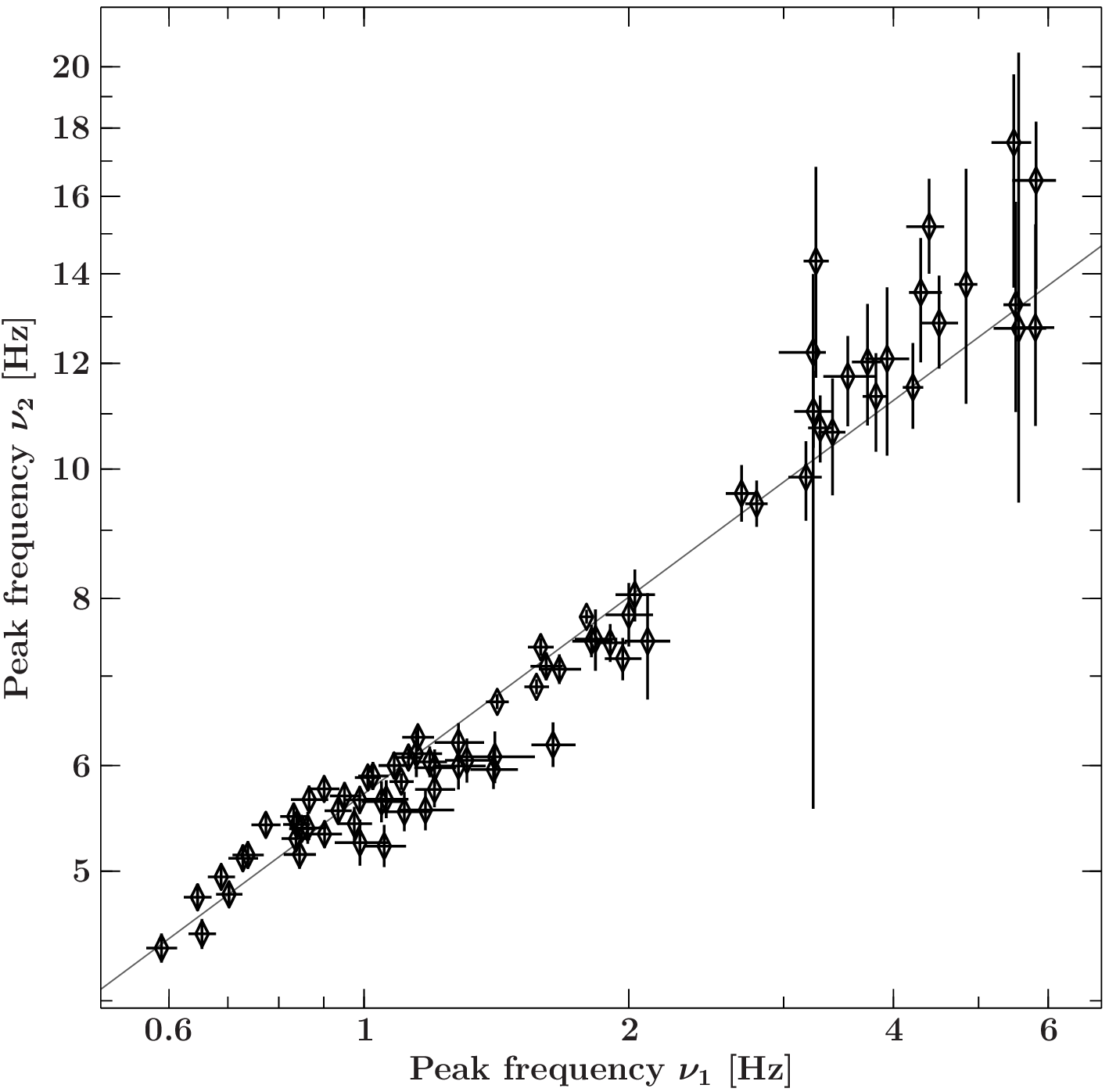}
  \caption{Correlation between the peak frequencies $\nu_1$ and $\nu_2$
    of the two Lorentzians.
    The gray line is the linear fit of Table~\ref{tab:v_g1_corr}.
  }
  \label{fig:L1_L2}
 \end{figure}
}

\newcommand\figGonePsd{
 \begin{figure*}\centering
  \includegraphics[width=.8\textwidth]{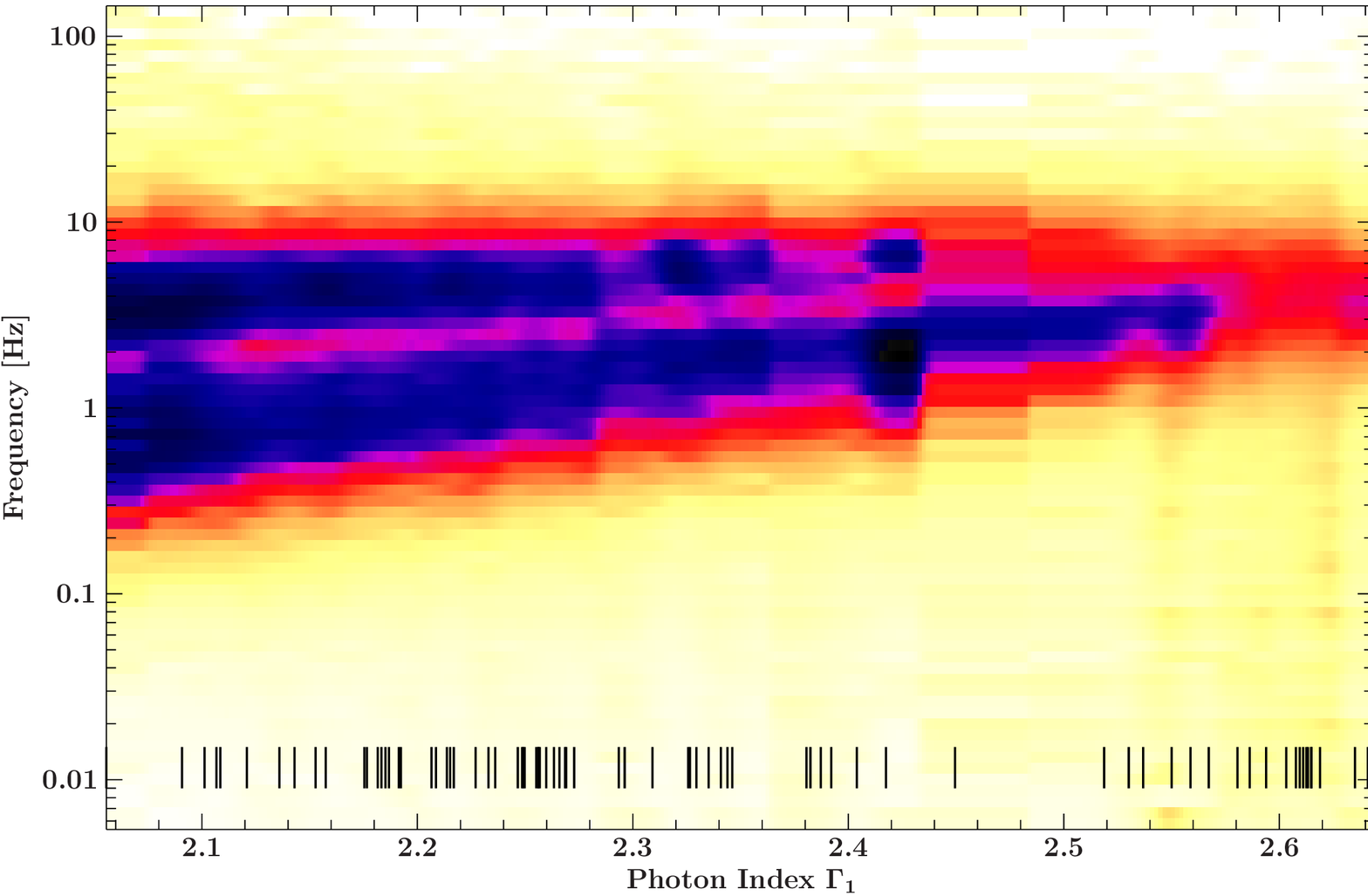}
  \caption{Variation of
    the 4.5--5.8\,keV power spectrum
    with the photon index $\Gamma_1$.
    A darker color code indicates a higher value of $f\times\text{PSD}(f)$.
    Dashes at the bottom show the photon indices of the individual spectra.
    Intermediate values are interpolated with a Gaussian blur.
    With increasing photon index the variability components shift to
    higher frequencies and their intensity decreases (in this energy
    band $L_2$ completely fades out toward $\Gamma_1\sim 2.6$).
  }
  \label{fig:g1_psd}
 \end{figure*}
}

\newcommand\figNuGoneCorr{
 \begin{figure}
  \includegraphics[width=\columnwidth]{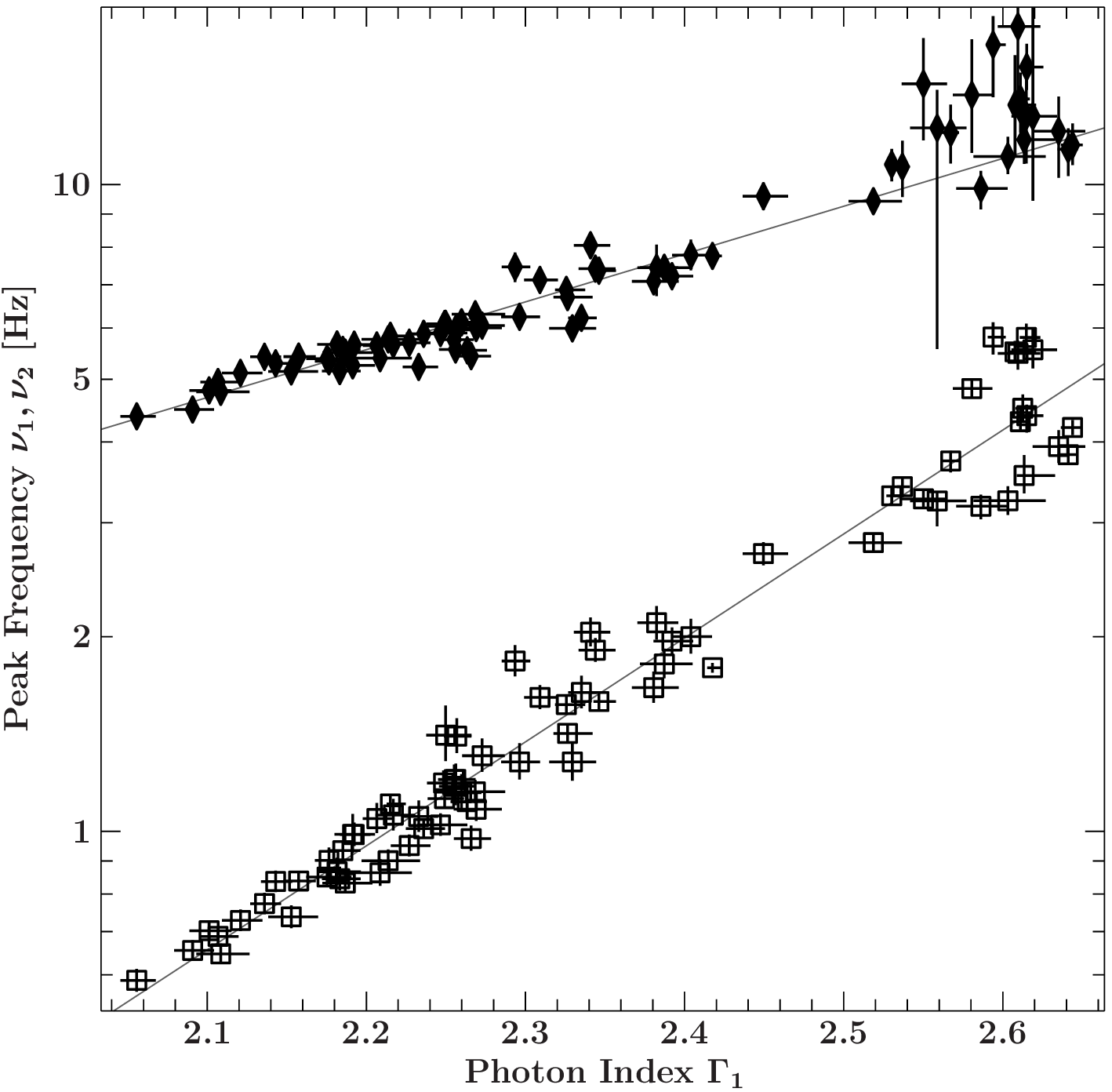}
  \caption{Correlation between the photon index and the peak frequencies
    of the Lorentzian profiles. The gray lines are the linear fits of
    Table~\ref{tab:v_g1_corr}.
  }
  \label{fig:v_g1_corr}
 \end{figure}
}

\newcommand\figPSDenergy{
 \begin{figure}
  \includegraphics[width=\columnwidth]{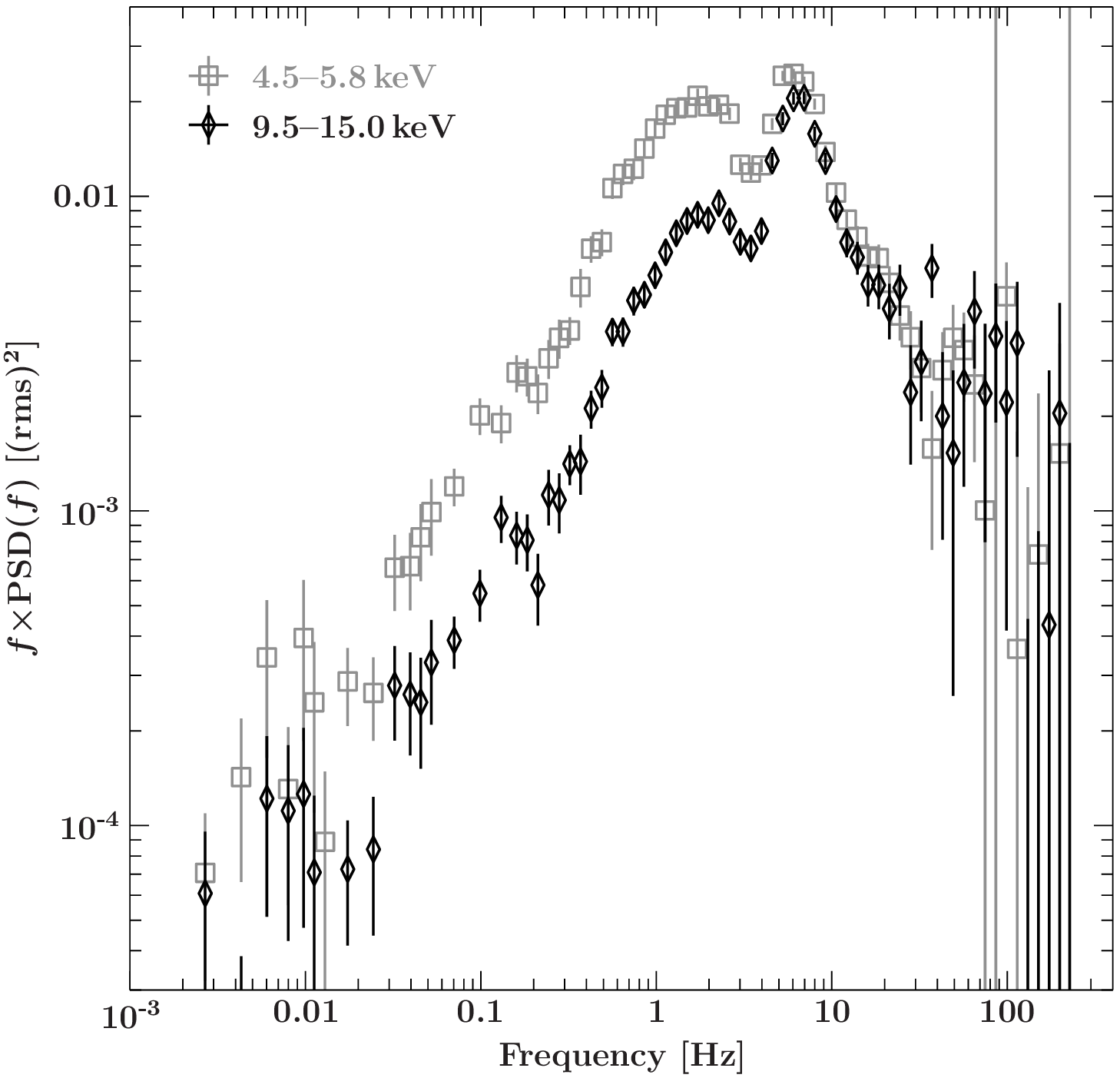}
  \caption{
    PSDs of the observation 90127-01-01-05 in the low (open squares) and
    high (black diamonds) energy band. The photon index of this
    observation is $\Gamma_1 = 2.33^{+0.02}_{-0.01}$.
  }
  \label{fig:PSD_energy}
 \end{figure}
}

\newcommand\figRmsEnergy{
 \begin{figure*}
  \includegraphics[width=\columnwidth]{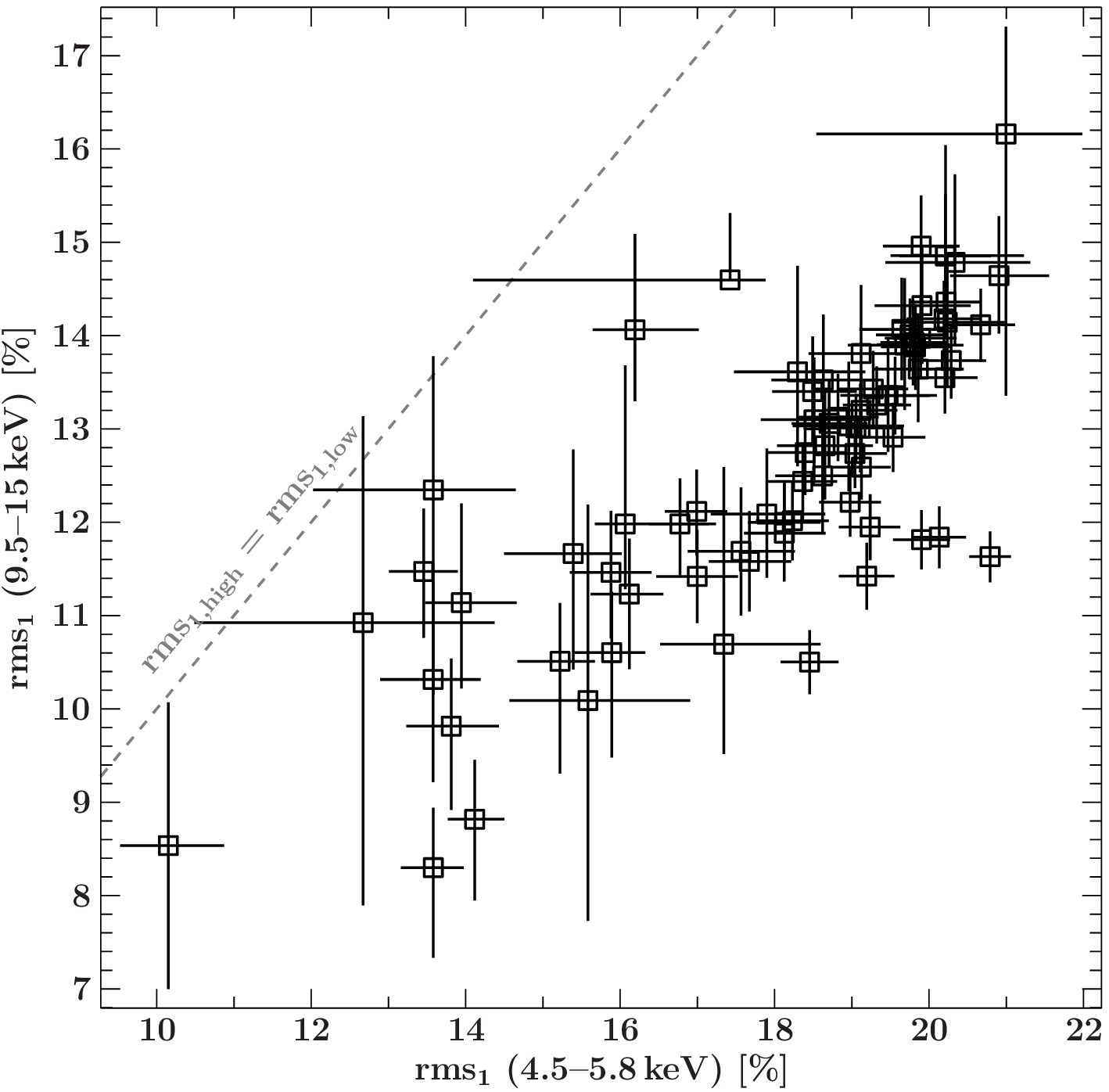}\hfill\includegraphics[width=\columnwidth]{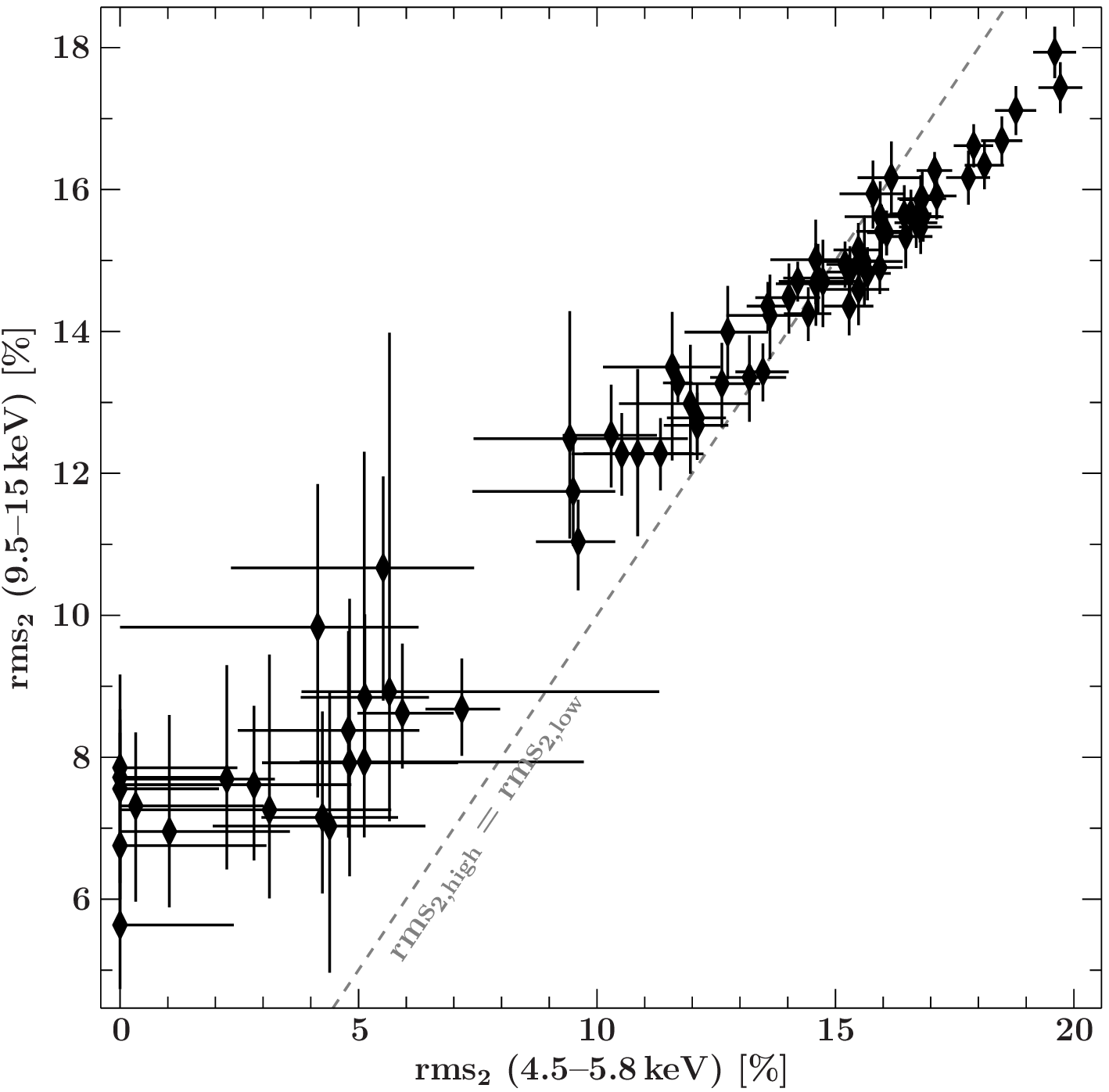}
  \caption{
    Strength of the Lorentzian profiles in the two energy bands,
    $\text{rms}_{i,\mathrm{high}}$ vs. $\text{rms}_{i,\mathrm{low}}$,
    for $L_1$ (left) and $L_2$ (right).
    The dashed line is the identity.
  }
  \label{fig:rms_energy}
 \end{figure*}
}

\newcommand\figMultiTiming{
 \begin{figure*}
  \includegraphics[width=\textwidth]{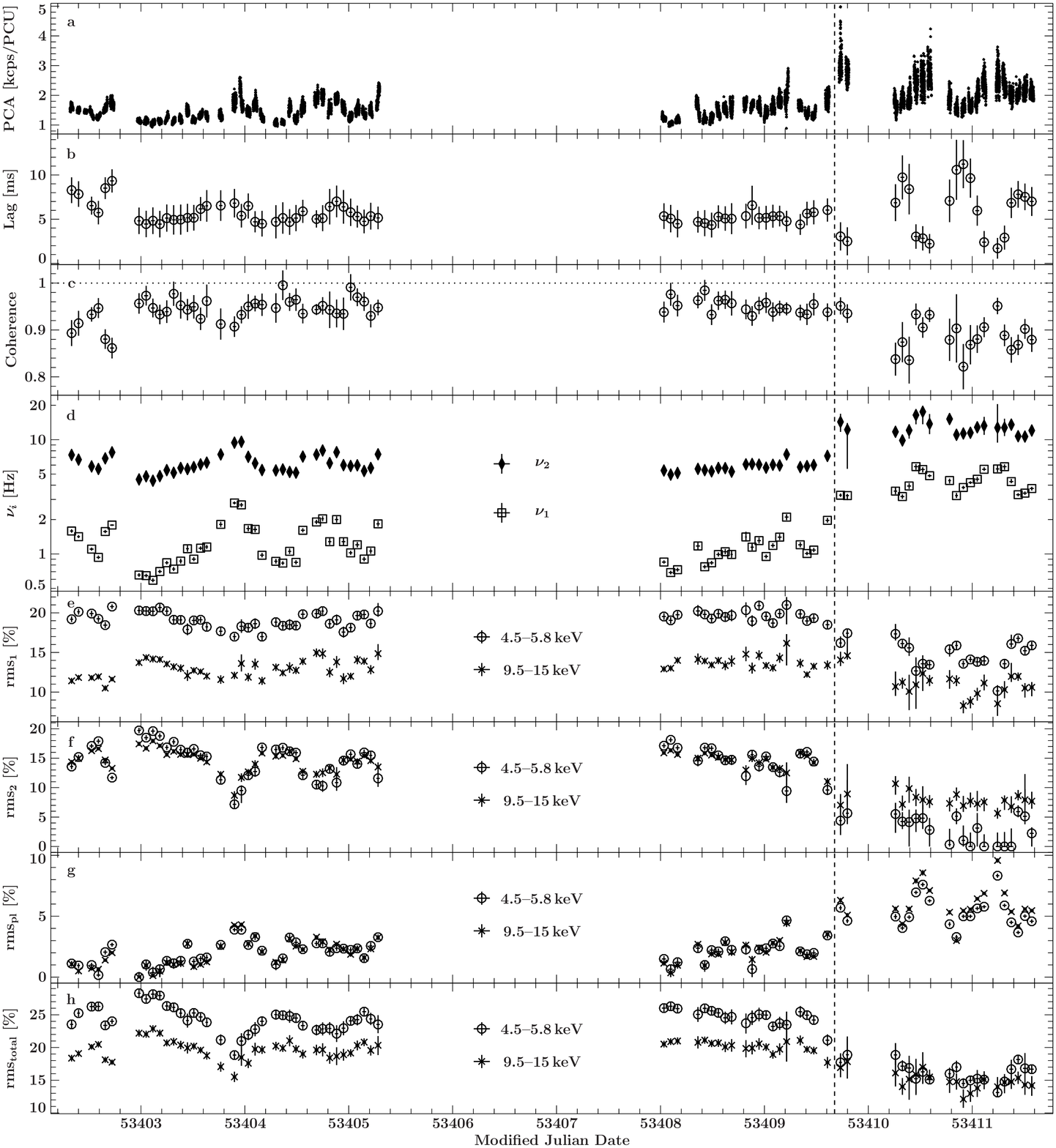}
  \caption{Evolution of the timing parameters during the observation.
    The upper panel (a) shows the PCA light curve of
    Fig.~\ref{fig:multi_spec}b for comparison,
    (b) the averaged time lag in the 3--10\,Hz band,
    (c) coherence in the 3--10\,Hz band, and
    (d) Lorentzian peak frequencies.
    The following panels show the rms in both energy bands (circles for
    the 4.5--5.8\,keV band and crosses for the 9.5--15\,keV band) of the
    components:
    (e) rms of $L_1$,
    (f) of $L_2$,
    (g) the power law, and
    (h) the total rms.
  }
  \label{fig:multi_timing}
 \end{figure*}
}

\newcommand\figHidLag{
 \begin{figure}
  \includegraphics[width=\columnwidth]{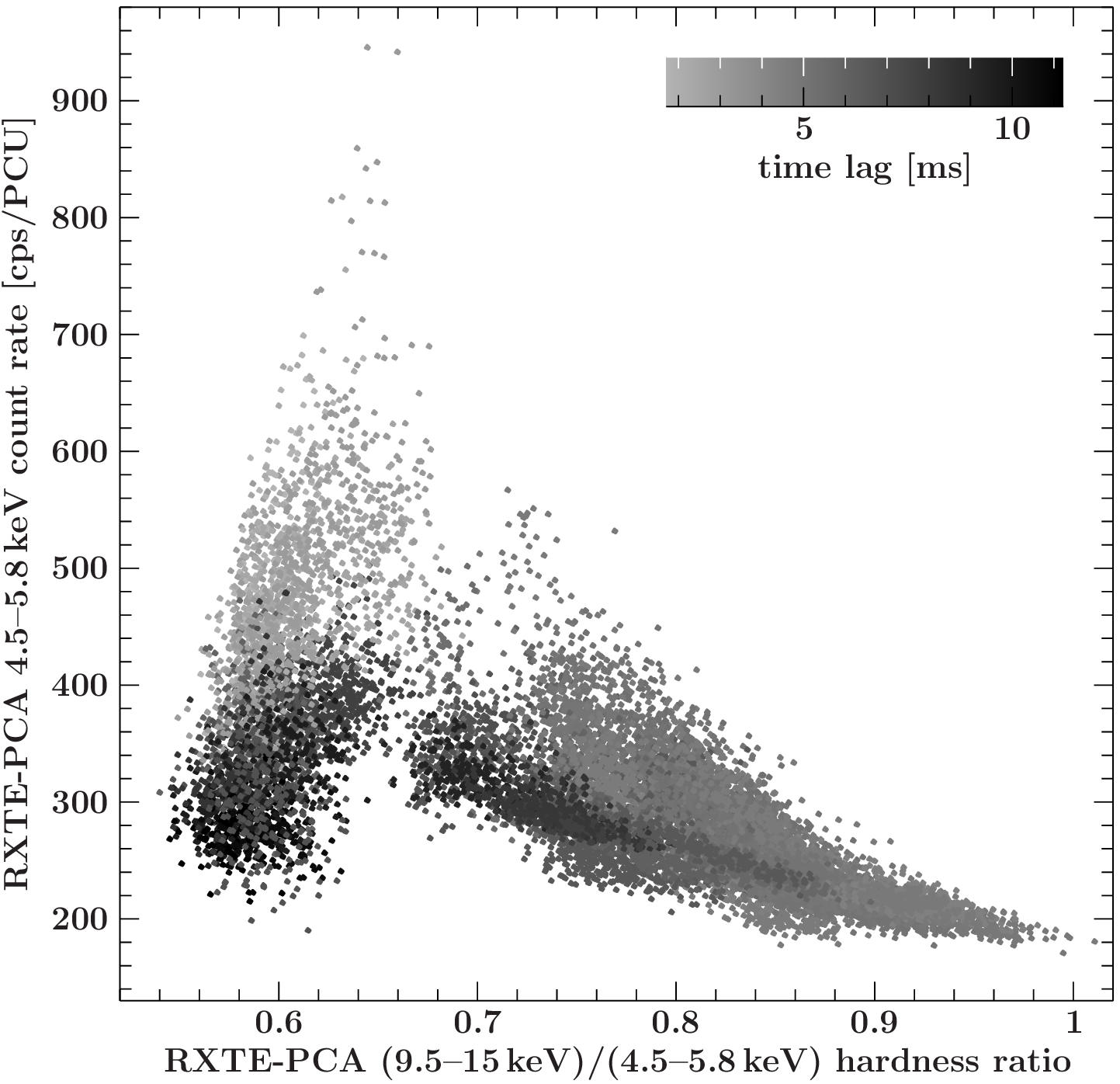}
  \caption{X-ray lag as a function of position in the hardness
    intensity diagram. See text for further
    explanation.
  }
  \label{fig:hid_lag}
 \end{figure}
}

\newcommand\figCplRms{
 \begin{figure}
  \includegraphics[width=\columnwidth]{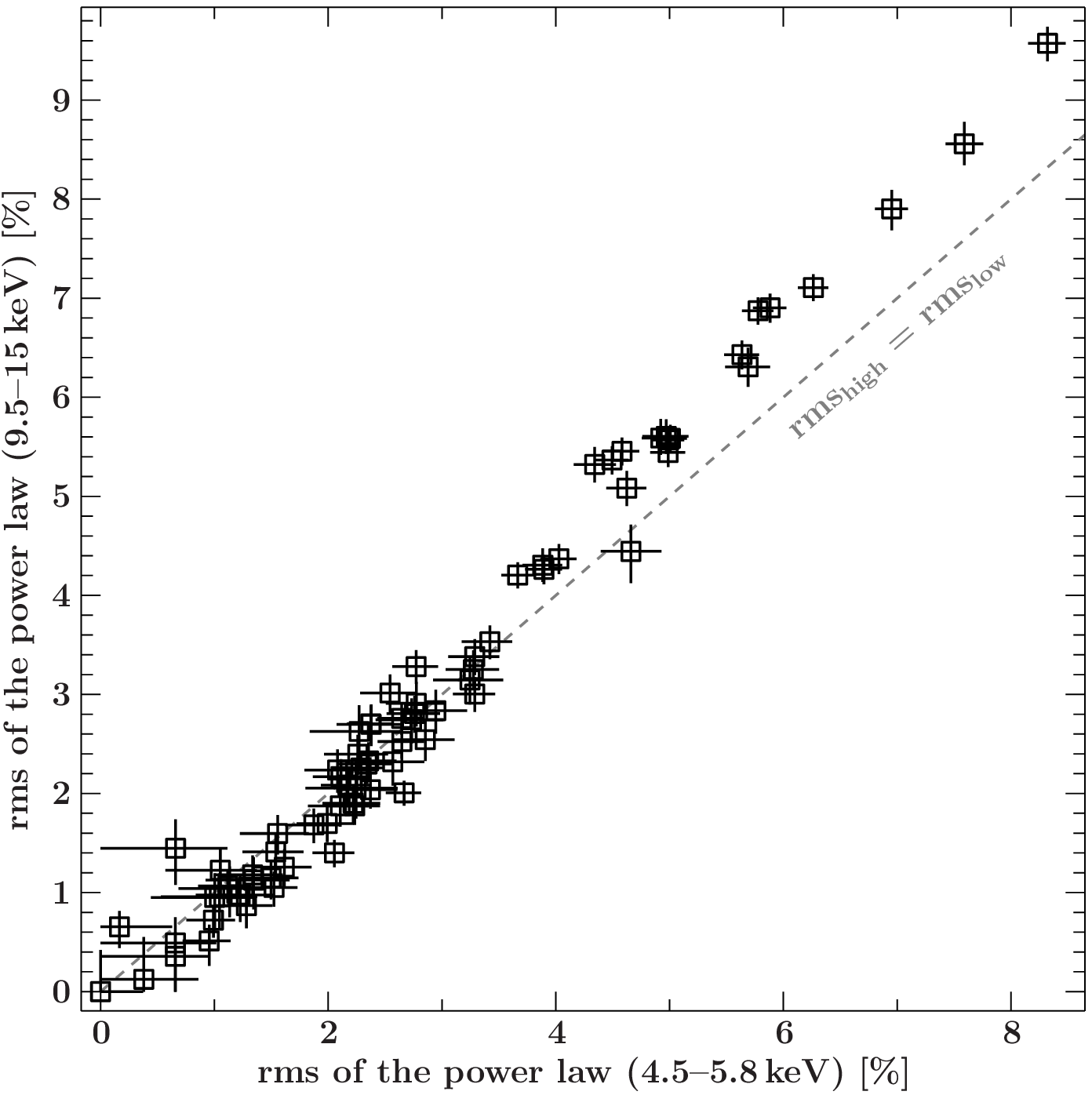}
  \caption{
    Contribution of the power law to the variability in the two energy
    bands. The dashed line is the identity.
  }
  \label{fig:cpl_rms}
 \end{figure}
}

\newcommand\figGoneQuality{
 \begin{figure}
  \includegraphics[width=\columnwidth]{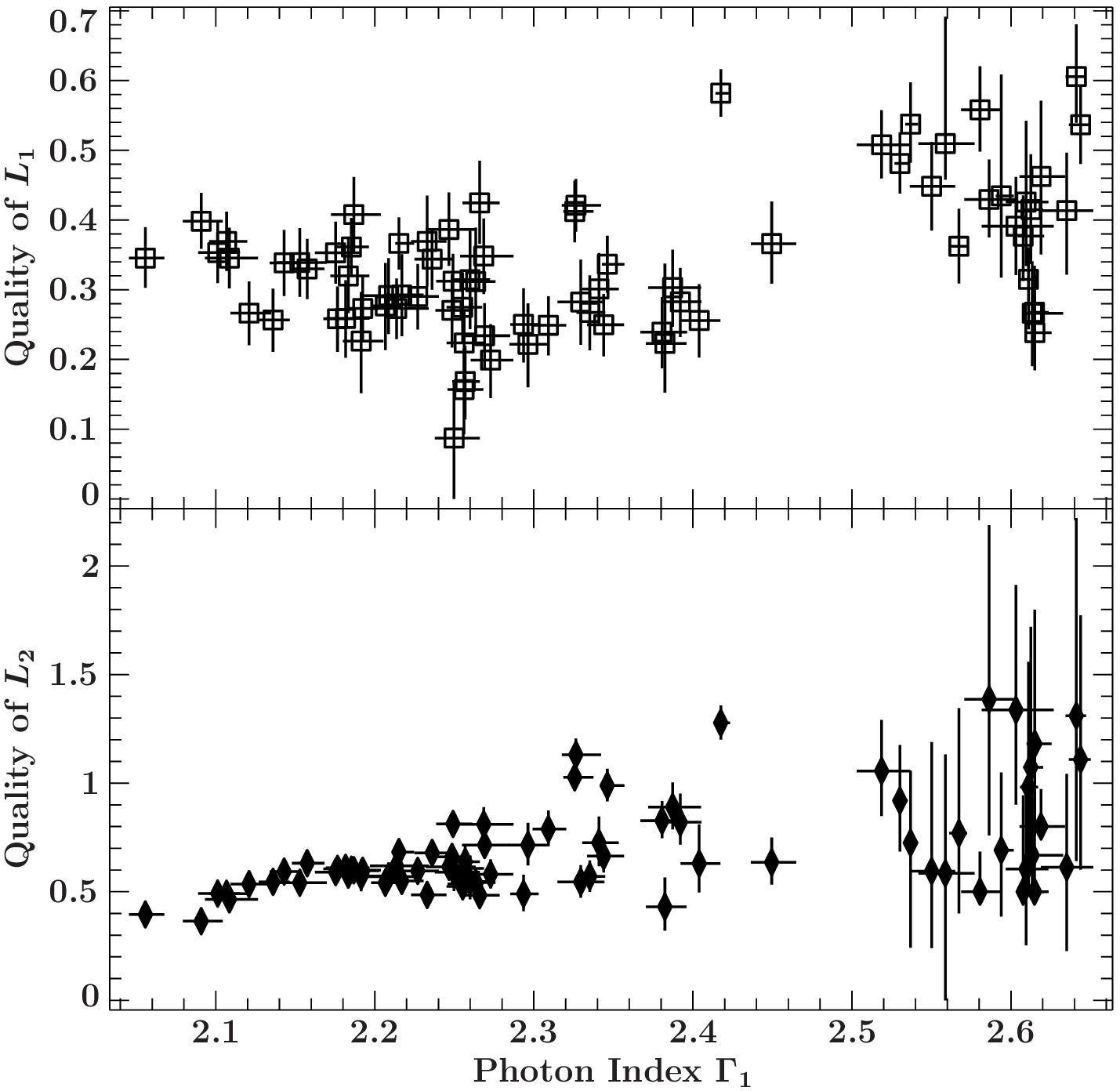}
  \caption{
    The quality factor of the Lorentzian profiles.
  }
  \label{fig:g1_quality}
 \end{figure}
}

\newcommand\figRms{
 \begin{figure*}
  \includegraphics[width=\columnwidth]{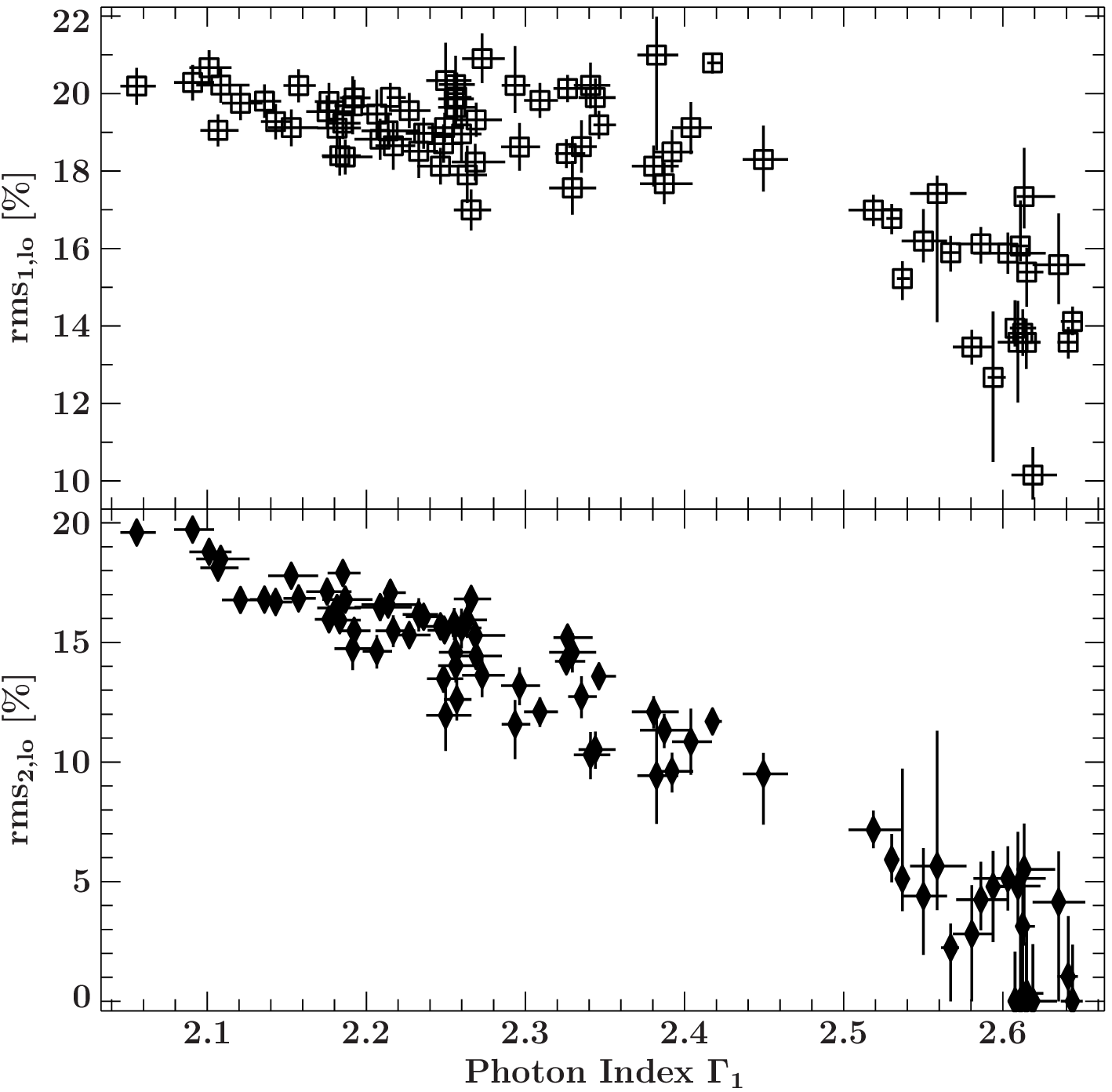}\hfill\includegraphics[width=\columnwidth]{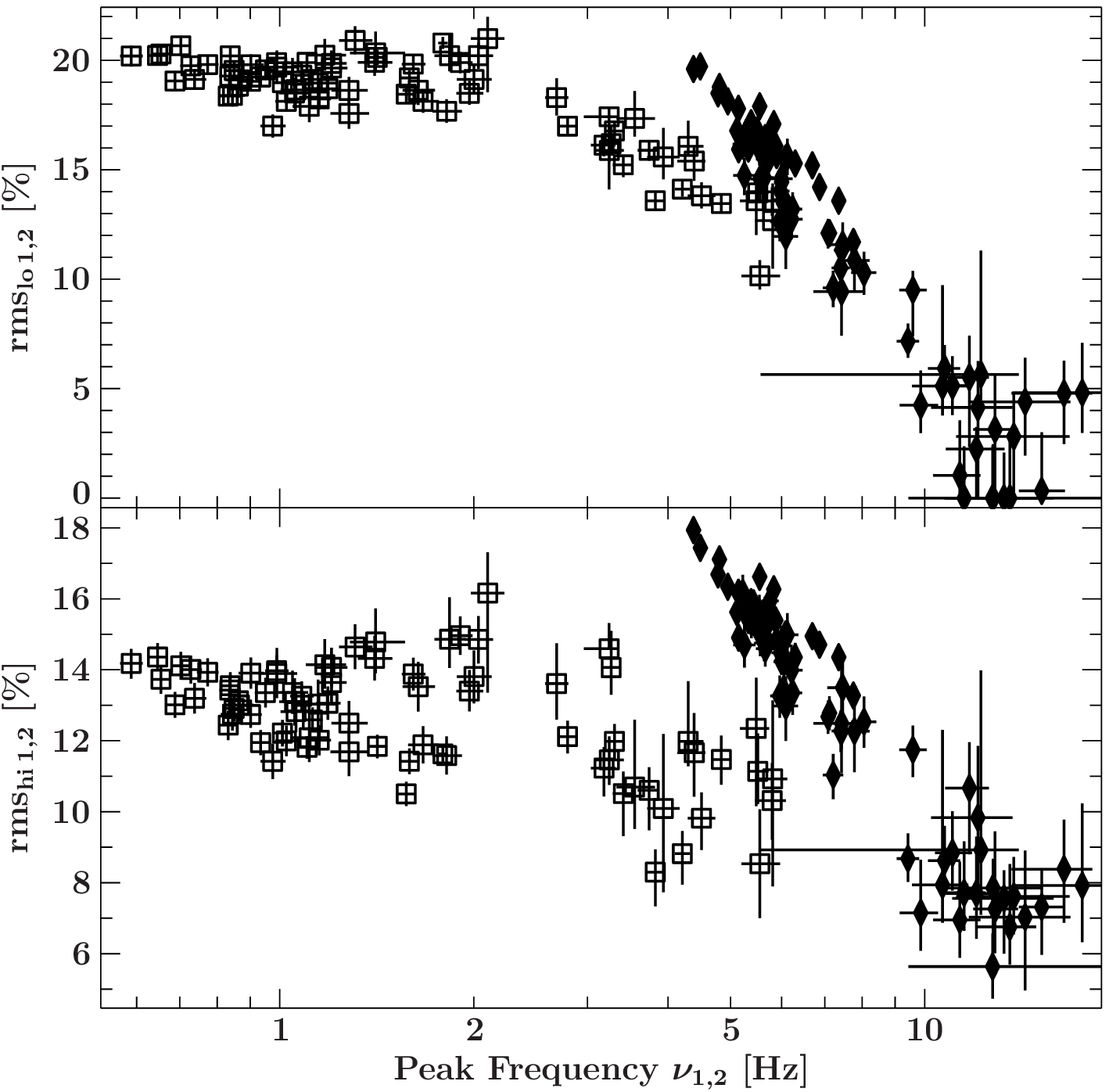}
  \caption{Dependence of the rms of the Lorentzian profiles on the
    photon index (left) and on the peak frequencies (right). The open
    squares correspond to the first Lorentzian and the filled diamonds to
    the second one.
  }
  \label{fig:rms}
 \end{figure*}
}

\newcommand\tabEnergyChannels{
 \begin{table}
  \caption{Energy bands of the \RXTE-PCA used in this paper according
    to the energy conversion of PCA Epoch 5.
  }
  \label{tab:energy_channels}
  \renewcommand{\arraystretch}{1.5}
  \begin{tabular}{lll}
   \hline
   PCA Channels & Energy (PCU0) & Energy (PCU1234)\\
   \hline
   11--13 &  4.53--5.82\,keV & 4.49--5.71\,keV\\
   23--35 &  9.72--15.4\,keV & 9.40--14.76\,keV\\
   \hline
  \end{tabular}
 \end{table}
}

\newcommand\tabNuGoneCorr{
 \begin{table}
  \caption{
    Correlation coefficients $\rho$ and weighted linear fits using the
    method of \citet{FasanoVio1988} between
    the Lorentzians' peak frequencies $\nu_i$ and the photon index
    $\Gamma_1$ (valid for $2.0 \leq \Gamma_1 \leq 2.7$).
  }
  \label{tab:v_g1_corr}
  \begin{tabular}{ll}
   \hline
   $\rho$ & linear correlation \Large\strut\\
   \hline
   0.98 & $\mathrm{ln}(\nu_{1}/\text{Hz}) = (-8.18 \pm 0.02)+(3.70\pm0.01)\cdot\Gamma_1$ \Large\strut\\
   0.96 & $\mathrm{ln}(\nu_{2}/\text{Hz}) = (-2.03 \pm 0.01)+(1.70\pm0.01)\cdot\Gamma_1$ \Large\strut\\
   0.98 & $\mathrm{ln}(\nu_1/\text{Hz}) = (1.743 \pm 0.007) + (0.489 \pm 0.017)\cdot\mathrm{ln}(\nu_2/\text{Hz})$ \Large\strut\\
   \hline
  \end{tabular} {\Large\strut}\\
 \end{table}
}

\begin{document}
\title{Spectro-timing analysis of Cygnus X-1 during a fast state transition}

\author{\mbox{M.~B\"ock\inst{1}} \and
  \mbox{V.~Grinberg\inst{1,2}} \and
  \mbox{K.~Pottschmidt\inst{3,4}} \and
  \mbox{M.~Hanke\inst{1}} \and 
  \mbox{M.~A.~Nowak\inst{5}} \and
  \mbox{S.~B.~Markoff\inst{6}} \and
  \mbox{P.~Uttley\inst{7}} \and  
  \mbox{J.~Rodriguez\inst{8}} \and  
  \mbox{G.~G.~Pooley\inst{9}} \and
  \mbox{S.~Suchy\inst{10}} \and
  \mbox{R.~E.~Rothschild\inst{10}} \and
  \mbox{J.~Wilms\inst{1}}
  }
\offprints{M.~B\"ock,\\ e-mail: {Moritz.Boeck@sternwarte.uni-erlangen.de}}
\institute{
Dr.\ Karl-Remeis-Sternwarte, Astronomisches Institut der
Universit\"at Erlangen-N\"urnberg, and Erlangen Centre for
Astroparticle Physics, Sternwartstra\ss{}e~7, 96049 
Bamberg, Germany
\and
Universit\"atssternwarte, Ludwig-Maximilians Universit\"at M\"unchen,
Scheinerstr.~1, 81679 Munich, Germany 
\and
CRESST, University of Maryland Baltimore County, 1000 Hilltop Circle,
Baltimore, MD 21250, USA 
\and
NASA Goddard Space Flight Center, Astrophysics Science Division, Code
661, Greenbelt, MD 20771, USA 
\and
MIT-CXC, NE80-6077, 77 Mass. Ave., Cambridge, MA 02139, USA
\and
Astronomical Institute ``Anton Pannekoek'', University of Amsterdam,
Kruislaan 403, Amsterdam, 1098 SJ, The Netherlands 
\and
University of Southampton, Southampton SO17 1BJ, UK
\and
Laboratoire AIM, CEA/IRFU - Universit´e Paris Diderot -
CNRS/INSU, CEA DSM/IRFU/SAp, Centre de Saclay, 91191 Gif-sur-Yvette,
France
\and
Cavendish Laboratory, J.~J.~Thomson Avenue, Cambridge CB3 0HE, UK
\and
Center for Astrophysics and Space Sciences, University of California
at San Diego, La Jolla, 9500 Gilman Drive, CA 92093-0424, USA 
}
\date {Accepted: 07/07/2011}

\abstract{We present the analysis of two long, quasi-uninterrupted
  \RXTE observations of Cygnus~X-1 that span several days within a
  10\,d interval.  
  The spectral characteristics during this observation cover the
  region where previous observations have shown the source to be most
  dynamic.  
  Despite that the source behavior on time scales of
  hours and days is remarkably similar to that on year time
  scales. This includes a variety of spectral/temporal correlations
  that previously had only been observed over \Cyg's long-term
  evolution. Furthermore, we observe a full transition from a hard to
  a soft spectral state that occurs within less than 2.5\,hours --
  shorter than previously reported for any other similar \Cyg
  transition.

  We describe the spectra with a phenomenological model dominated by a
  broken power law, and we fit the X-ray variability power spectra
  with a combination of a cutoff power law and Lorentzian components.
  The spectral and timing properties are correlated: the power
  spectrum Lorentzian components have an energy-dependent amplitude,
  and their peak frequencies increase with photon spectral index.
  Averaged over 3.2--10\,Hz, the time lag between the variability in
  the 4.5--5.7\,keV and 9.5--15\,keV bands increases with decreasing
  hardness when the variability is dominated by the Lorentzian
  components during the hard state. The lag is small when there is a
  large power law noise contribution, shortly after the transition to
  the soft state.  Interestingly, the soft state not only shows the
  shortest lags, but also the longest lags when the spectrum is at its
  softest and faintest.  We discuss our results in terms of emission
  models for black hole binaries.}


\maketitle

\section{Introduction}
\label{sect:intro}
The observed behavior of most black hole (BH) X-ray binaries (XRBs)
can be classified into two major states, the so-called hard, power-law
dominated, and the soft, thermally dominated state \citep[e.g.,][and
references therein; these authors also introduce the most extreme
steep power law state seen in only a few
BHs]{RemillardMcClintock2006}. In the hard state 
the X-ray continuum can be described by a hard power law, which is
interpreted as resulting from Comptonization in a hot
($kT_\mathrm{e}\approx50$--100\,keV) electron plasma surrounding the
accretion disk, or in a jet
\citep{Dove1998,HaardtMaraschi1993,Zdziarski2002,markoff:05a}.  
A strong contribution of jet synchrotron emission is found in some
sources \citep{Markoff2001,Russell2010,Laurent2011}. In the case of
\Cyg the X-ray continuum can be modeled with a broken power law hardening
above a break energy $E_\text{break}\approx10$\,keV and a high-energy
cutoff with a folding energy of $\sim$100\,keV
\citep[e.g.,][]{Wilms2006}. In the hard state, BH XRBs are generally
observable in the radio with a flat/inverted radio through IR spectrum
\citep{FenderBelloniGallo2004,Fender2009,Dunn2010}.
X-ray variability above 0.1\,Hz during this state is high
($\mathrm{rms}>20\%$).
In the soft state
the X-ray spectrum is dominated by emission from an accretion disk and
the core radio/IR emission is quenched. \Cyg, however, retains a strong
power law component even in its softest X-ray spectra.

In most Galactic BHs, transitions between states are common.
The state behavior
generally can be described as part of a characteristic evolution of
spectral and timing states that depends on the previous source
history. Over the outburst of an X-ray transient, the source is seen
to move along a \textsf{q}-shaped track in a hardness intensity
diagram \citep[HID; e.g.,][]{FenderBelloniGallo2004,HomanBelloni2005}.
The outburst starts in the hard state; during the initial phase,
the source brightens without significantly changing its spectral shape,
and the radio/IR flux increases. 
At some point, the X-ray spectrum starts to soften.
The radio/IR flux becomes more erratic.  In some sources radio ejections
were seen during this intermediate
state
\citep{Corbel2004,FenderBelloniGallo2004,Fender2006,Wilms2007}. After
the state transition, the source spectrum is soft and no core radio/IR
emission is detected \citep{Fender2009}. During this soft state phase
the bolometric source flux decreases, until at some lower luminosity
the X-ray spectrum hardens, and the source becomes detectable in the
radio/IR again. During the final part of the outburst the source is back
in the hard state and the source luminosity decreases until the XRB
ceases to be observable
\citep{FenderBelloniGallo2004,Belloni2006}. 

The observed behavior of BH XRBs is determined by the interplay
between emission from the accretion disk and from the regions of the
accretion inflow/outflow in which the hard X-rays and the radio emission
are generated. Transitions between the two states are therefore
especially suited to place constraints on the physics of these
components.  Despite the fact that the outbursts can last weeks or
months, the transitions between hard and soft states are for most
sources rare and often fast. 
\citet{Belloni2006}, for example, report a hard to soft transition of
\mbox{GX\,339$-$4} that occurred in less than 10\,h. For this reason,
the behavior during state transitions is not well studied. In
addition, comparisons between different sources are complicated because
that the influence of other variables, such as the variation of
the mass accretion rate, is difficult to ascertain. Here we will
present a detailed analysis of a state transition as observed in a
single object.

The canonical BH candidate \object{Cygnus X-1} is a persistent XRB
that spends most of the time in the hard state; however, it shows
frequent excursions into soft states \citep[][and references
  therein]{Wilms2006}, including aborted transitions \citep[``failed
  state transitions'';][]{Pottschmidt2000,Pottschmidt2003}.  Because
of its brightness and persistent emission, \Cyg is one of the best
studied BHs.  In addition to studies of snapshot observations
\citep[e.g.,][]{gierlinski:99a,disalvo:01a,frontera:01a,makishima:08a,hanke:09a},
most recent work has concentrated on the long-term behavior of the
source from a monitoring campaign with the \textsl{Rossi X-Ray
  Timing Explorer} (\RXTE) at a two week sampling that
started in 1998 and is still ongoing
\citep[][and references
  therein]{Pottschmidt2003,gleissner:04a,gleissner:03a,Axelsson2005,Wilms2006}.
The motivation of the present work was to fill the gap between the
biweekly time scale and individual \RXTE pointings with an intensive
quasi-continuous observation spanning 10\,days.  Our aim was to
determine differences and similarities between the behavior of \Cyg on
time scales of minutes, hours, days, and the long-term behavior found
during the monitoring campaign. As we will show in the following, this
observation covered an especially interesting time when a state
transition occurred.

The remainder of this paper is structured as follows: In
Sect.~\ref{sec:data} we describe the observations and data analysis
methods, the results of which are presented in
Sects.~\ref{sec:analysis} and \ref{sec:timing_analysis}.
The results are discussed in Sect.~\ref{sec:summary}.

\section{Observations and data analysis}\label{sec:data}
This paper is based on 78 \RXTE orbits of observations of Cygnus~X-1
on 2005 February 01--11 (MJD\,53402--53412; \RXTE proposal ID 90127).
We used data from all three instruments onboard \RXTE,
namely the All Sky Monitor
\citep[ASM;][]{Levine1996}, the Proportional Counter Array
\citep[PCA;][]{Jahoda2006} and the High Energy X-ray Timing Experiment
\citep[HEXTE;][]{Rothschild1998}. We reduced data from PCA and HEXTE
with \textsc{HEASOFT}~6.3.1 following \citet{Wilms2006}, 
but using data from all xenon layers and making use of
the improvement of background models, which now allow us to
consider data taken at least 10\,minutes after passages through the
South Atlantic Anomaly (SAA; see \citealt{Fuerst2009}).  Spectra and
lightcurves with a time resolution of 16\,s were obtained using the
standard2f mode PCA data.  To characterize the source's spectral shape
on this short time scale using hardness studies, light curves were
extracted for the two channels listed in
Table~\ref{tab:energy_channels}, as well as for the whole energy range
of the PCA. For the timing analysis light curves with a time
resolution of $\sim$2\,ms ($2^{-9}$\,s) were extracted in the same
energy channels. The calculation of Fourier-frequency dependent
quantities such as power spectra, X-ray time lags, and the coherence
function follows \citet{Nowak1999timing}. Data modeling and all other
data reduction was performed with ISIS 1.4.9
\citep{Houck2000,Noble2006,NobleNowak2008}.

\tabEnergyChannels

To study the spectral variability of the source, we extracted spectra
for each \RXTE orbit, yielding 78 spectra with a PCA exposure of about
3\,ks each.
We considered PCA data between 2.8\,keV and 50\,keV, and HEXTE data in
the energy range 18--250\,keV. For the fourth and fifth PCA bin
(2.8--3.2\,keV and 3.2--3.6\,keV) we added systematic errors of 1\%
and 0.5\%, respectively.\footnote{
  Based on modeling all Cyg X-1 spectra from the years-long RXTE
  monitoring campaign we found this selection for the two low-energy
  bins to provide reliable absorption and disk parameters. We did not
  add the recommended systematic errors of 0.5\% to the remaining PCA
  bins because the reduced $\chi^2$ of fits to hard state observations of \Cyg
  can be significantly below 1 in this case, indicating an
  overestimation of the uncertainties.}
A multiplicative constant was used to  
account for the different PCA and HEXTE flux calibration. The constant
for the HEXTE had typical values of about 85\%
with respect to the PCA.

Hard X-ray spectra of BHs are traditionally explained using
Comptonization, either in a Compton corona or in a jet \citep[][and
  references therein]{Wilms2006,markoff:05a}. A disadvantage of these
models, however, is that the spectral parameters obtained from these
models are often not unique \citep{nowak:08b,Nowak2011}.  It is still
unclear which of the several models used in the literature is the
correct physical interpretation of the data.  Following
\citet{Wilms2006}, we therefore modeled the X-ray spectra using a purely
empirical spectral model, an absorbed broken power law with
exponential cutoff.
In the following, we will denote
the power law indices below and above the break energy
$E_\text{break}\approx10$\,keV with $\Gamma_1$ and $\Gamma_2$,
respectively. A Gaussian profile was added to model the Fe
K$\alpha$-line at 6.4\,keV. For the thermal emission of the accretion
disk the \texttt{diskbb} model \citep{Mitsuda1984} was applied.
This phenomenological model with 
$\Gamma_1>\Gamma_2$ describes the continuum with its spectral
hardening above 10\,keV very well, with reduced $\chi^2$ values around
1.
We note that it is possible to map a spectrum defined through
$\Gamma_1$ and $\Gamma_2$ to the more physical models, using the
correlations found by \citet{Wilms2006}. Clear
correlations between $\Gamma_1$ and the other parameters of the
empirical model allow us to identify the X-ray spectral state of the
source with just one parameter, simplifying the analysis of the state
dependence of X-ray timing and radio properties.

\figHid

In addition to the X-ray data, quasi-simultaneous radio observations
were made at 15\,GHz with the University of Cambridge's Ryle
Telescope. The time resolution was 8\,s, using interleaved
observations of a nearby phase reference source and observations of
3C48 and 3C286 as flux density calibrators. The telescope uses
linearly-polarized feeds, and the measurements represent Stokes' I+Q;
the flux density scale is believed to be consistent with the scale of
\citet{Baars1977}, after allowing for the difference in polarizations.
The overall amplitude scale is believed to be accurate to about 5\%,
and the noise level on a single 8-s sample is about 12\,mJy (the data
shown in Fig.~\ref{fig:multi_spec} have been averaged over 10\,min,
and so the thermal noise is about 1.4\,mJy).

\section{X-ray spectral variability}\label{sec:analysis}

\subsection{An hour-long state transition}

To characterize the evolution of the source's overall spectral
properties, in Fig.~\ref{fig:hid} we show the HID of \Cyg as obtained
from the single-dwell 90\,s \textsl{RXTE}-ASM data taken between 1996
and 2010.  In this diagram, the hard state is found at a hardness of
$>$1 and at count rates $\lesssim$40\,cps.
Transitions into the soft state are characterized in the ASM by a
softening of the spectrum correlated with a brightness increase,
followed by a further softening and a brightness decrease, leading to
a $\Lambda$-like shape in this HID. The right panel of
Fig.~\ref{fig:hid} shows a similar HID, calculated from the
\textsl{RXTE}-PCA 16\,s lightcurves during the 2005 February
observation only.  The figure shows that at this time resolution a
$\Lambda$-shaped transitional region between hard and soft state in
the HID is covered. In the following we will consider the two regions
covered in the HID obtained with \textsl{RXTE}-PCA as a ``hard'' and
``soft'' region instead of ``hard intermediate'' and ``soft
intermediate'' regions.

In order to investigate the potential occurrence of a state
transition in more detail we now turn to the spectral parameters of the broken
power-law modeling of our orbit-resolved PCA and HEXTE spectra. The
X-ray and radio lightcurves and the evolution of the spectral
parameters are shown in Fig.~\ref{fig:multi_spec}.

\figMultiSpec

During the 10 days of the 2005 February observation, we find $2.0 <
\Gamma_1 < 2.7$. For broken power-law models, 
\citet{Wilms2006} show that $\Gamma_1>2.1$ is consistent with
transitional and soft states.  Thus the range of $\Gamma_1$-values
during our observation confirms the conclusion from the HID that \Cyg
spent these days close to the transitional region between the hard and
the soft state. A closer inspection of Fig.~\ref{fig:multi_spec} shows
that at MJD\,53409.7, from one \RXTE pointing to the next, \Cyg
changed its photon index from $\Gamma_1{=}2.39$ to $\Gamma_1{=}2.55$
within only 2.25\,hours (Fig.~\ref{fig:multi_spec}, panel~e) and moved
in the HID from the harder region to the softer region. No other
transition between the two regions was included in these data.
The change is also seen in the more prominent disk, whose peak temperature
increased to 0.25--0.40\,keV and whose inner radius moved inward to
a mean value of $17\pm3\,r_\mathrm{G}$ from $56\pm5\,r_\mathrm{G}$
before the transition.\footnote{Values of $r_\mathrm{in}$ were
obtained  from the \texttt{diskbb} normalization assuming a distance
of 2\,kpc, a black hole mass of $10\,M_\odot$, and an inclination of
$35^\circ$ \citep{Herrero1995}.}
A drop in $N_\mathrm{H}$ after the transition could be caused by
increased ionization of the stellar wind by the much softer X-ray
spectrum. The duration of the transition was only a few hours, much
shorter than the day-long transitions previously
reported for \Cyg \citep[e.g.,][]{cui:96a,cui:01a} and comparable with
the very short transition seen in GX~339$-$4 by \citet{Belloni2006}.

\subsection{Spectral variability on time scales from hours to days}

In contrast to previous observations, which consisted either of a few
hour-long pointed observations or of several ks long observations
spaced by weeks, the present data set also allows us to study the
spectral variability of \Cyg during a transitional state.
Figure~\ref{fig:G1_G2} compares the variability of the two photon
indices, $\Gamma_1$ and $\Gamma_2$, during the 2005 February
observation, with that observed in the long-term monitoring between
1999 and 2004 \citep{Wilms2006}. During the 10\,d of our observation,
the variation of $\Gamma_1$ and $\Gamma_2$ covered $\sim$33\% of the
parameter space covered by the long-term monitoring, where
$1.6<\Gamma_1<3.4$ was found.

\figGoneGtwo

This general similarity between the monitoring on time scales of hours
and weeks also extends to the broad-band radio to X-ray spectral
variability. The relation between the \RXTE orbit-wise
determined X-ray photon index $\Gamma_1$ and the simultaneously
measured 15\,GHz radio flux, was consistent with the relation yielded
by complete observations in the long-term monitoring campaign
(Fig.~\ref{fig:g1_radio}). In the hourly \emph{and} in the weekly
monitoring we observe that the radio flux is seen to increase while
the X-ray spectrum softens up to $\Gamma_1\approx2.2$, and the radio
lightcurve shows flaring behavior. Above this threshold, i.e., during the
transition into the soft state, the radio flux decreases. 

These results indicate that the overall spectral variability of \Cyg
on time scales of hours is similar to that seen on time scales of
months. The large scatter
seen from one monitoring observation to the next is due to the fact
that the variability of the properties of the source on time scales of
hours covers a significant fraction of the overall variability of the
source seen in the X-ray monitoring.
Because the radio and X-ray fluxes also show similar behavior down
to time scales of hours, the region where the radio and X-ray flux are
produced must be small.

\figGoneRadio

We expect the tight correlation between the X-rays and the radio to
break down on time scales of minutes, especially during flaring
episodes, which are interpreted as ejections of hot electron bubbles
from the accretion flow into the jet
\citep{vanderlaan:66a,hjellming:88a,gleissner:04a}. Because the radio emission is
produced farther downstream in the jet, a delay
between the radio and X-rays is expected during these events, as was 
indeed seen in the
one flare reported by \citet{Wilms2007} for \Cyg and is also
observed in other microquasars, in particular in GRS\,1915$+$105
\citep{Mirabel1998, Klein-Wolt2002, Prat2010}, where it was
interpreted as the ejection of the corona \citep{Rodriguez2008}.
The most likely flaring
event during our observation occurred directly after the state
transition, when the PCA light curve in its total energy range of
about 2--60\,keV peaked to
$5000\,\text{counts}\,\text{PCU}^{-1}$, one of the highest measured
\Cyg count rates ever observed with \RXTE. This count rate was
reached although the absorption was high with a hydrogen
column of $3.6^{+0.6}_{-0.5}10^{22}\,$cm$^{-2}$. Unfortunately, however,
no simultaneous radio data are available during this time.

\section{X-ray timing behavior}\label{sec:timing_analysis} 

Having discussed the spectral variability on time scales of ks, we
now turn to the behavior of \Cyg on time scales of seconds. We will
first characterize this behavior through standard power spectrum
analysis in several energy bands, and then discuss the variability
through Fourier cross-spectral 
analysis techniques, namely X-ray
time lags and the coherence function.

\figPsdSeq

\subsection{Power spectra}\label{sec:psds}
The calculation of the power spectral density (PSD) was performed
following \citet{Nowak1999timing} and \citet{Pottschmidt2003}. Power
spectra are normalized according to \citet{Miyamoto1991}, i.e.
\begin{equation}
 \int \mathrm{PSD}(f)\;{\rm d}f = \left(\frac{\sigma}{\mu}\right)^2 \;\;,
\end{equation}
where $\mu$ is the mean value of the light curve and $\sigma$ its
standard deviation.
The PSDs were modeled as the sum of a power law with an exponential cutoff
and two Lorentzian profiles $L_1$ and~$L_2$.
We parameterize $L_i$ by
the peak frequency $\nu_i$ where $f\times L_i(f)$ reaches its maximum,
the quality factor $Q_i=f_i/\Delta f_{i,\textnormal{FWHM}}$,
and the strength rms$_i$, where
$\text{rms}_i^2 = \int_0^\infty L_i(f)\;{\rm d}f$
corresponds to the contribution to the relative root mean square variability.
Using the peak frequency as independent parameter instead of the resonance frequency
\begin{equation}
 f_i \;\;=\;\; 2\,Q_i\,\nu_i\,(1+4Q_i^2)^{-1/2}
 \label{eq:fi}
\end{equation}
facilitates the fitting procedure, because $\nu_i$ is much better
constrained and less dependent on $Q_i$ than $f_i$, especially for
broad Lorentzians with low $Q_i$.  $\nu_i$ is also the relevant
parameter for correlations \citep[e.g.,][and references
  therein]{Pottschmidt2003}.  $L_i$ has thus the form
\begin{equation}\label{eqn:L}
 L_i(f) \;\;=\;\; A_i \, \left(\nu_i^2-2f_i f + f^2\right)^{-1} \;,
\end{equation}
where $f_i$ is defined by Eq.~(\ref{eq:fi}) and
\begin{equation}
 A_i \;\;=\;\; \text{rms}_i^2\,\nu_i\,\left(1+4Q_i^2\right)^{-1/2}\left[\frac{\pi}{2} +\arctan(2Q_i)\right]^{-1} \;.
\end{equation}
This parameterization naturally comprises ``zero frequency-centered Lorentzians''
\citep[e.g.,][]{Nowak2000}
for $Q_i=0$ (and thus $f_i=0$), but $\nu_i>0$.

\figLoneLtwo

\figGonePsd

\figNuGoneCorr

\tabNuGoneCorr

In our earlier work based on data from 1998 to 2001, we required up to
four Lorentzians to model the typical power spectrum
\citep{Pottschmidt2003}. For the data analyzed here, two broad
Lorentzians in addition to a power law with an exponential cutoff
describe the PSDs accurately enough \citep[see also][]{Axelsson2005}.
The major reason for this difference is that the 1998--2001 data
include observations with harder spectra than the data analyzed here
(see also below and Fig.~\ref{fig:psd_seq}).
Because a strongly rising power law with strong cutoff can affect the rms
of broad Lorentzians, we froze the index of the cutoff power law to
1. In combination with the Lorentzian profiles, this $1/f$
``flicker noise'' describes the data well, but it does not extend to
the highest observed frequencies and a cutoff to the power law was
required.

\figPSDenergy

\subsubsection{Correlation with the X-ray spectrum}\label{sec:psdavg}

During the observation the peak frequency of $L_1$, $\nu_1$, varied
between 0.5 and 6\,Hz, while the peak frequency of $L_2$ varied
between 4 and 30\,Hz. These peak frequencies are clearly correlated
(Fig.~\ref{fig:L1_L2}); interestingly, we find $\nu_1 \propto \sqrt{\nu_2}$
(see Table~\ref{tab:v_g1_corr}).
Figure~\ref{fig:psd_seq} displays typical power spectra from the
hardest to the softest observations. The shape of the power spectra
clearly depends on the spectral state, and exhibits the following
features:

\figRmsEnergy

\begin{itemize}
\item The PSDs of observations with a hard X-ray spectrum show a clear
  double humped profile. Individual Lorentzians can be clearly
  identified. A few of the hardest observations in this sample
  ($\Gamma_1\lesssim2.1$)
  show slight indications of the presence of a third
  broad Lorentzian with a peak frequency in the range of 30--60\,Hz.
  The contribution of the low frequencies to the variability of the
  corresponding light curve is very small, i.e., $f\times\text{PSD}(f)$
  decreases strongly toward low frequencies and no power law
  component is apparent.
\item With softening X-ray spectrum, the low-frequency
  ($\lesssim0.1$\,Hz) variability
  increases. In this frequency range the PSD has a power law shape.
  The Lorentzian components shift to higher frequencies.
\item In the spectrally softest observations the power law with
  exponential cutoff dominates the PSD. The rms of the second
  Lorentzian $L_2$, which is located at higher frequencies, is
  significantly reduced with respect to $L_1$.
\end{itemize}

Figure~\ref{fig:g1_psd} shows how the power spectrum changes as a
function of the photon index $\Gamma_1$. The image shows the humps of the PSDs as
dark bands, which become weaker when the X-ray spectrum gets softer. In
addition it is apparent that the peak frequencies of the Lorentzians
are clearly correlated with the photon index $\Gamma_1$ (see also
Fig.~\ref{fig:v_g1_corr}). See Table~\ref{tab:v_g1_corr} for the
numerical relation between the Lorentzian parameters and the spectral
shape.

\subsubsection{Energy dependence of the
  variability}\label{sec:energy_dependence} 

Figure~\ref{fig:PSD_energy} shows the shapes of two power spectra
calculated for the same observation but in different energy bands.
While the qualitative shape of the PSDs is similar and the
corresponding peak frequencies are found to agree within their
uncertainties, the strength of the individual components of the PSD is
clearly energy-dependent. The most obvious disparity is the strength
of the humps.
This energy dependency can be quantified 
using the Lorentzian model.
The peak frequencies and quality factors of the individual Lorentzians
are generally energy-band-independent within their statistical
uncertainty. The rms of the Lorentzians, however, is strongly
energy-band-dependent.

We fitted the power spectra from both energy bands simultaneously, 
constraining $\nu_i$ and $Q_i$ to have the same value in all energy
bands and leaving the Lorentzian rms a function of the energy
band. This approach allows us to 
localize the Lorentzians in the few cases when they are almost faded
out in one energy band but still visible in the other one.
Figure~\ref{fig:rms_energy} shows the energy dependence of the rms for
both Lorentzians.

The strength of $L_1$ in the 9.5--15\,keV energy band is only about two thirds
of that in the 4.5--5.8\,keV band. 
With an rms variability
of $\gtrsim$10\%, $L_1$ was visible in all the low-energy data
analyzed here.
The highest rms values for $L_2$ are found in the observations with
the hardest X-ray spectra (see Figs.~\ref{fig:g1_psd} and
\ref{fig:rms_energy}).  At these high values, the rms for $L_2$ is
greater in the 4.5--5.8\,keV band than in the 9.5--15\,keV band.  As
the spectrum softens, $\text{rms}_2$ decreases in both energy bands,
but decreases more significantly in the 4.5--5.8\,keV energy band. In the
softest observations, the $L_2$ component cannot be detected in the
4.5--5.8\,keV band, while it remains visible in the 9.5--15\,keV
energy band with an rms value of a few per cent
(Fig.~\ref{fig:rms_energy}, right).
In the soft state a strong power law component is observed at the
frequencies where the $L_1$ component dominates during
the hard state.

\figMultiTiming

\subsection{Time lags and coherence function}
While light curves at different energies are generally correlated,
this relationship is frequency-dependent. The degree of
linear correlation between the two light curves can be measured with
the coherence function, which depends on the Fourier frequencies
\citep[a detailed discussion is given by][\S~3.1]{Nowak1999timing}.
For two correlated time series one can define a Fourier frequency-dependent delay or time lag
$\delta_t(f_j)$ \citep{nowak:96a} as
\begin{equation}\label{eq:lag}
\delta_t (f_j) = \frac{\arg[S^{\ast}(f_j) H(f_j)]}{2\pi f_j}
               = \frac{\Phi_{H(f_j)}-\Phi_{S(f_j)}}{2\pi f_j}.
\end{equation}
Here $S(f_j)$ and $H(f_j)$ are the discrete Fourier transformations of the
soft and hard X-ray lightcurves $s(t_i)$ and $h(t_i)$,
and $\Phi_{S(f_j)}$ and $\Phi_{H(f_j)}$ are the corresponding
phases. In Eq.~\ref{eq:lag}, the phase was chosen such that a positive time lag
means that the light curve in the hard energy band is delayed with
respect to the soft energy band, as is common for BH XRBs.
A detailed description of the calculation of time lags and coherence,
including their interpretation and the calculation of their
uncertainties, is given by \citet{nowak:96a}. 

The lag is generally a strong function of Fourier frequency. To obtain
a simpler quantity, we
averaged the time lag spectrum and the coherence function in the
3.2--10\,Hz band \citep[see][]{Pottschmidt2003}.  We use these
averaged coherence and time lag values in the following.  Their
evolution is shown in Fig.~\ref{fig:multi_timing}.  In this data set
the coherence was close to 1 and the time lag was usually longer than
$\sim$4\,ms. After the transition to the soft state there are
observations in which the time lag drops to about 2\,ms. The time lags
vary strongly between about 2 and 11\,ms after the transition, and the
coherence changes rapidly. This behavior is flux-dependent: While
short time lags occur at the highest count rates, the time lags are
long during the soft state observations with relatively low count
rates.

Figure~\ref{fig:hid_lag} again shows the HID of the present
observation.
The X-ray hardness and intensity in Fig.~\ref{fig:hid_lag} are
calculated on a resolution of 16\,s, while the X-ray lags, indicated
in the figure by grayscale, have been calculated on a time scale of
individual \RXTE orbits, which typically had an exposure of 3\,ks.
Different regions in the HID are dominated by different characteristic
lag values. In the hard state the time lag is shortest at the largest
hardness and increases as the X-ray spectrum softens. In the soft state the
X-ray time lag is short at the highest count rates, while
it is strongly enhanced at low count rates.

The shortest hard state lag found (about 4\,ms) is above the lowest
values found by \citet{Pottschmidt2003}, which is consistent with the
fact that there are only observations with $\Gamma_1\gtrsim2.0$ in our
data set. Observations with harder X-ray spectra and shorter lags are
not included in the sample analyzed here.

Previous work indicated that the time lags in typical soft states were
quite short (2--3\,ms; comparable to the shortest lags seen in hard
states), and that the time lags increased during transitional states
\citep{Pottschmidt2003}. 
The temporal evolution shown in Fig.~\ref{fig:multi_timing} reveals
that the time lags are short (2--3\,ms) when the variability
contribution of the power law noise is large, whereas the Lorentzian
variabilty close to the transition is associated with long time lags
(4--11\,ms depending on spectral hardness).
This is consistent with the results of \citet{Pottschmidt2003}, where
the shortest time lags in the soft state arose from variability power
spectra dominated by a cutoff power law, as opposed to Lorentzian
shape.

\section{Discussion and summary}\label{sec:summary}

\subsection{Discussion}

\subsubsection{State transition}

The changes of the emission properties of \Cyg are continuous. Because the
source does not follow the complete \textsf{q}-shaped track in a
hardness intensity diagram, which is found in outbursts of X-ray
transients \citep{FenderBelloniGallo2004,HomanBelloni2005,Dunn2010}, it is
difficult to specify well-defined borders between its different
emission states.  We find the following changes of properties: In the
classical hard state of \Cyg strong radio emission is observed, which
is thought to originate from a steady jet. The intensity of the radio
emission increases as the spectrum softens up to about $\Gamma_1
\approx 2.1$. Above this value, when \Cyg leaves its classical hard
state \citep{Wilms2006},
the radio flux decreases in correlation with the hard X-ray flux
\citep{gleissner:04a}. 

In the transitional state data analyzed here, we see an additional
change of the emission properties at $\Gamma_1 \approx 2.5$, where the
HID separates and the behavior of the time lags changes.  Remarkably,
this result is indicated by both spectral and timing analysis. The
data points fall into two clearly separated regions in the HID,
also corresponding to different power spectral shapes.
Transient BH XRBs only drastically change their timing 
properties during transitions from ``hard intermediate states'' to
``soft intermediate states''.  The fractional rms drops because of the
disappearance of flat-top noise components (described as Lorentzians
in this paper) in favor of a $1/f$ (power law) noise (Miyamoto et
al., 1993; Takizawa et al., 1997; Nespoli et al., 2003; Homan \&
Belloni, 2005; Casella et al., 2004; Casella et al., 2005; Belloni et
al., 2005; Belloni, 2009).  In addition, the ``soft intermediate
state'' power spectra may contain specific QPOs referred to as type-A
and type-B (Casella et al., 2005).

\citet{Nespoli2003} found a QPO at about 6\,Hz in an observation of
GX\,339$-$4 during a transition from the hard to the soft state. The
frequency of the QPO was correlated to the PCA count rate. The power
spectra on short time scales (16\,s) of the data analyzed here did not
show a comparable QPO, but it indicated 
that the peak frequencies of the Lorentzian profiles are correlated
with the PCA count rate even on time scales shorter than one
\textsl{RXTE} orbit. During the hard state the PCA count rate is
anticorrelated with the spectral hardness. Similar to \Cyg,
GX\,339$-$4 shows increasing time lags as the soft count rate
increases in the hard state and the flat-top noise disappears when the
GX\,339$-$4 softens into a ``soft intermediate state'' \citep{Belloni2005}.

\figHidLag

\subsubsection{Noise components}

Our model consisting of only two Lorentzian profiles in addition to a
power law with exponential cutoff is clearly justified by the
structure of the power spectra shown in Fig.~\ref{fig:g1_psd}.
A third Lorentzian is only slightly
indicated in a few of our hardest observations \citep[consistent with][]{Pottschmidt2003},
but is not required in general.  Our identification of the Lorentzian
profiles is equal to that of \citet[see also the discussion in their
  Sect.~4.1]{Axelsson2005}, and our results are consistent with
theirs: The $\nu_1$--$\nu_2$ relation is identical,
and it is also visible here that rms$_2$ decreases uniformly with the
peak frequency $\nu_2$.

Power spectra in the two energy bands used here can be modeled
simultaneously with the normalization as the only energy-dependent
parameter. This means that a single physical process 
is responsible for producing the individual variability components,
but that the contribution of each process to the variability is
energy-dependent.

\figCplRms

There are different physical models describing the origin and the
behavior of the noise components observed in the power spectra. Some
models connect the observed frequencies with fundamental general
relativistic ones of perturbed orbits near the compact object
\citep[see][for a summary]{NowakLehr1998}. Among others, mechanisms
discussed are perturbed epicyclic frequencies, the Lense-Thirring
precession frequency, and ``diskoseismic'' modes that are related to
maximum epicyclic frequency. Our observed peak frequencies,
0.5--30\,Hz,
correspond to epicyclic frequencies at about 22--345 gravitational
radii for a $10M_\odot$ compact object.  \citet{NowakLehr1998} point
out that the ``diskoseismic'' frequencies are very sensitive to
luminosity changes, have a low rms variability, and thus cannot be
applied to systems with $\geq$10\% rms
as in our observation.  In addition, it is difficult to explain the
spectral hardness of the variability components.
Compared to the Lorentzian profile $L_1$, the one at high
frequencies, $L_2$, is more dominant in 
the high-energy band, which is clearly above energies at which the
disk emission contributes strongly, thus it is unlikely that both
variabilities originate from the same region.
\citet{Wilkinson2009} observe a low-frequency Lorentzian profile
originating from the accretion disk. \citet{Uttley2011} point out that the
disk can produce high-frequency variations as well, which would not
show up as strong soft variations, provided that they occur at small
disk radii so that only a small fraction of disk flux is modulated as
they propagate inward. In contrast to $L_1$, the absolute variability
contributions of $L_2$, and also that of the $1/f$ noise component, are
comparable in both energy bands, however.

\citet{Psaltis2000} postulate a transition radius in the inner disk
acting as a low band-pass filter, which yields a dynamical model for
the production of modulations, with a strong resonance to orbital,
periastron- and nodal-precession. The model accounts for the observed
PSDs and explains correlations between the characteristic frequencies.
We find $\nu_1 \propto \sqrt{\nu_2}$, which cannot be identified
clearly with one of the correlations between the modes calculated by
\citet{Psaltis2000}. An extrapolation of this relation implies
a characteristic frequency of $\sim$30\,Hz, at which $\nu_1 =
\nu_2$.

\figGoneQuality

In the soft state the variability (rms) at high frequencies is lower,
indicating a relatively steady state within the analyzed time scales.
A possible interpretation of the power law arising in the PSDs as the
spectrum softens is flicker noise in the accretion disk
\citep{Lyubarskii1997}. The data analyzed here indicate, however, that
the variability contribution of the $1/f$ component is strong in both
energy bands and slightly stronger in the high-energy band compared to
the low-energy band in the cases of strong $1/f$ noise contribution,
i.e., when the X-ray spectrum is soft (Fig.~\ref{fig:cpl_rms}).
A detailed interpretation of the soft state power spectra is presented
by \citet{Misra2008}, who consider linear and non-linear responses of
a damped harmonic oscillator, 
which is assumed to govern the disk variability at each radius.

\figRms

The dependence of the Lorentzians' rms on their peak frequency
(Fig.~\ref{fig:rms}) is similar to that presented by
\citet{Pottschmidt2003} in Figure 6 of their work. There is, however,
one slight difference: Their Figure 6 indicates a
unique relation between the peak frequency and the rms of all
Lorentzians below $\sim$20\,Hz, i.e., when one Lorentzian shifted to a
frequency at which another Lorentzian used to be, its rms was the
same as the other Lorentzian had at this frequency. This relation
could be interpreted as a self-similarity, i.e., one Lorentzian
profile replaces another one when it shifts to its frequency and
originates from the same process or the same
mode. Figure~\ref{fig:rms} does not show this behavior.  Owing to the
different energy dependence of the Lorentzians it is obvious that this
relation depends on the considered energy channels. It is nevertheless
possible that one Lorentzian replaces another one at its frequency,
because when L$_1$ shifts to the position of L$_2$ other properties of
the source also change. For example $\nu_2$ is about 5\,Hz for
$\Gamma_1\approx2.0$, whereas $\nu_1$ reaches this frequency at
$\Gamma_1\approx2.7$. This change in the X-ray spectrum implies
changes in the emitting region, possibly affecting the properties of
the variability components.

\subsubsection{Time lags}

Previous time lag studies of \Cyg showed typically short lags around
2--3\,ms in the hard state, while enhanced lags up to 9\,ms were
observed during ``failed state transitions''. \citet{Pottschmidt2003}
used the energy bands 2--4\,keV and 8--13\,keV to obtain these values.
When the source entered deeper into the soft state, the time lag
returned to short values \citep{Pottschmidt2000}.

Our new data set with higher temporal resolution (\RXTE orbit-wise
extractions) allows us to extend these results
(Fig.~\ref{fig:multi_timing}, \ref{fig:hid_lag}).
In the hard state the
time lag increases with softening X-ray spectrum consistent with
previous observations, whereas after the transition to the soft state
the time lag behavior changes. Depending on the brightness, the time
lag changes from values of up to 11\,ms at low count rates to values
of 2--3\,ms at high count rates without appreciable softening of the
source.
Changes from long to short time lags occur from one \RXTE orbit to
the next. 
Figure~\ref{fig:multi_timing} shows that the lag is short when the
variability contribution of the power law is large, whereas the time
lags of the Lorentzian variability are long depending on their peak
frequency and thus on the spectral hardness. 

Time lags can originate from reprocessing of radiation and propagation
in the emitting region.
\citet{Nowak1999compton} exclude disk-intrinsic time delays as the
sole reason for the lag, and discuss constraints on properties of the
corona if the time lag originates in it.
\citet{Kylafis2008} present a jet model that can reproduce the
behavior of the time lags in the \Cyg hard state.
\citet{KoerdingFalcke2004} show that time lags can be caused by
a pivoting power law spectrum arising, e.g., from a jet/synchrotron
model. In their work
the sign of the time lag depends on the hardness/flux correlation
\citep[][Fig.~9]{KoerdingFalcke2004}. 
Interestingly, the correlation visible in the lambda-shaped HIDs seems
to change its sign coinciding with the change in time lag behavior,
exactly at the transition:
in the hard state the count rate increases with decreasing hardness,
while it is the other way around in the SIMS.
The transition is also where the behavior of the time lag changes.

\citet{Uttley2011} show that the time lags seen at soft X-rays
(0.5--9\,keV) are caused by the accretion disk. They point out that
Comptonization models for lags have difficulties, whereas propagation
delays \citep[see, e.g.,][]{Kotov2001,Arevalo2006} can explain their
data, and that a similar mechanism can cause time lags at higher energies.

The clear correlations between spectral and timing properties put
strong constraints on models explaining the emission of accreting
black holes.

\subsection{Summary}
The main results of this paper can be summarized as follows:
\begin{itemize}
\item \Cyg was in an intermediate/soft state during
  2005 February 01--11. The photon index $\Gamma_1$
  reached values between 2.0 and 2.7 and was correlated with $\Gamma_2$ in the
  same way as found by \citet{Wilms2006}.
\item The source was very variable during that period of time; the
  total PCA count rate varied between 1000 and 5000 counts$\rm\:s^{-1}$ per PCU.
\item A full transition from the hard state 
  to the soft state
  occurred in less than 2.5\,hours. The two states
  can be clearly distinguished in a hardness intensity diagram.
\item Spectral and timing parameters are strongly correlated.
  The peak frequencies of the Lorentzian profiles shift to
  higher values with increasing $\Gamma_1$.
  Those
  are identical in the low- and the high-energy band,
  whereas their rms
  depends on the energy.
\item In the HIMS the time lag increases with decreasing
  hardness. After the transition to the SIMS
  the time lag behaves differently. At high count rates
  (${\gtrsim}2500\,\mathrm{counts}\:\mathrm{s}^{-1}\,\mathrm{PCU}^{-1}$
  in this case) the time lag drops to about 2\,ms and the contribution of
  the power law noise is large. At lower count rates the time lag is
  long, with
  values up to 11\,ms.
\end{itemize}

\acknowledgements{We acknowledge funding from the
\textsl{Bundes\-ministerium f\"ur Wirtschaft und Technologie} through
the \textsl{Deutsches Zentrum f\"ur Luft- und Raumfahrt} under
contract 50OR0701.
M.A.N. was supported by NASA Grant SV3-73016.
J.R. acknowledges partial funding from the European Community's
Seventh Framework Programme (FP7/2007--2013) under grant agreement
number ITN 215212 ``Black Hole Universe''.
We thank the International Space Science Institute, Berne, Switzerland
for supporting team 116 and the X-ray group at the Center for
Astrophysics and Space Sciences of the University of California at San
Diego for their hospitality during the initial stages of this project.
}

\end{document}